\documentclass[aps,prd,twocolumn,groupedaddress,nofootinbib,showpacs]%
{revtex4-1}

\usepackage[T1]{fontenc} 
\usepackage{bm}
\usepackage{amsmath}
\usepackage{amssymb}
\usepackage{graphicx}
\newcommand{\thorn}{\text{\TH}}
\newcommand{\omicron}{{o}}

\newcommand{\bk}{{\bm k}}
\newcommand{\bn}{{\bm n}}
\newcommand{\bak}{{\acute{\bm{k}}}}
\newcommand{\bx}{{\bm x}}

\begin{document}

\title{Fermions and gravitational gyrotropy}

\author{Adam D. Helfer}

\email[]{helfera@missouri.edu}
\affiliation{Department of Mathematics and Department of Physics \& Astronomy,
University of Missouri,
Columbia, MO 65211, U.S.A.}

\date{\today}

\begin{abstract}
In conventional general relativity without torsion, high-frequency
gravitational waves couple to the chiral number density of spin one-half quanta:  the polarization of the waves is rotated by $2\pi N_5 {\ell _{\rm Pl}^2}$, where $N_5$ is the chiral column density and $\ell _{\rm Pl}$ is the Planck length.  This means that if a primordial distribution of gravitational waves with E-E or B-B correlations passed through a chiral density of fermions in the very early Universe, an E-B correlation will be generated.  This in turn will give rise to E-B and T-B correlations in the cosmic microwave background (CMB).
Less obviously but more primitively, the condition Albrecht called ``cosmic coherence'' would be violated, changing the restrictions on the class of admissible cosmological gravitational waves.  This altered class of waves would, generally speaking, probe earlier physics than do the conventional waves; their effects on the CMB would be most pronounced for low ($\lesssim 100$) multipoles.  Rough estimates  indicate that if the tensor-to-scalar ratio is less than about $10^{-2}$, it will be hard to constrain a spatially homogeneous primordial $N_5$ by present data.
\end{abstract}



\maketitle

\section{Introduction}
\label{sec:intro}

Two, presumably deeply interfused, puzzles in cosmology are the origin of leptons and the emergence of the chiral character of their spectrum and interactions which is familiar at ordinary scales.  These must develop in the very early Universe, but just how early is unclear.
We do not know the physical mechanisms involved, and the only model-independent constraint on the energy-scales is that they must be beyond those which have been substantially explored in collider experiments.  
In view of this, it is worthwhile considering general properties of fermions in cosmological contexts.  

I describe here some remarkable effects involving fermions and gravitational waves, which should on ordinary scales be tiny but could be significant in cosmology.  These effects grow out of a striking coupling of gravitational radiation to chiral physics:  a chiral current $j^5_a$ (of right- minus left-handed spin one-half particles) rotates the polarization of high-frequency waves by an angle
\begin{equation}\label{phrot}
  2\pi \ell _{\rm Pl}^2 N_5
\end{equation}
in the left-handed sense, where $\ell _{\rm Pl}$ is the Planck length 
(we take $\ell _{\rm Pl} =\sqrt{G}$, with $G$ Newton's constant, 
and Planck's constant $\hbar$ and the speed of light $c$ to be unity)
and
\begin{equation}\label{ccd}
  N_5=\int _\gamma j^5_a\, d\gamma ^a
\end{equation}
is the {\em chiral column density} along the null geodesic $\gamma$ giving the geometric-optics trajectory of the wave.    (The rotation is measured relative to parallel transport; it is not a gauge-dependent effect.)
This rotation may be called {\em gravitational gyrotropy}, in parallel with the corresponding optical effect.\footnote{In particle physics, the quantity $j^5_a$ is usually called the axial current, ``chiral current'' being used for a related but different quantity.  However, in the present context, ``chiral current,'' and especially ``chiral column density,'' are more precise and descriptive terms than their ``axial'' alternatives would be.}

As we will see, the Planck-area prefactor in eq. (\ref{phrot}) makes direct observation of the rotation in the present Universe unlikely, but a gyrotropy in the very early Universe could leave detectable relics.  If (as is widely supposed) gravitational waves were created primordially, they would be affected by passage through a subsequent chiral current density.  This effect is potentially quite significant, because it violates a constraint (``cosmic coherence'') which is usually assumed.  
These changes would in turn modify the waves' effect of the cosmic microwave background (CMB).  
The tight-coupling regime would be more deeply probed.
Besides quantitative effects, there would be the creation of 
opposite-parity correlations (T-B or E-B correlations, with T temperature, E ``electric'' polarization and B ``magnetic'' polarization), which are forbidden in the standard models.

\subsection{Contrast with other effects}

This effect is distinct from others which have been considered:

First, it is a prediction of conventional general relativity, without the addition of any torsion term or other modification.  While there is a long history (beginning with Kibble \cite{Kibble1961} and Sciama \cite{Sciama1964}) of suggesting that fermions should alter Einstein's theory through the introduction of torsion, or of introducing explicitly chiral gravitational terms (typically higher-derivative, Chern--Simons like terms, e.g. \cite{LWK}), here no change at all is made in the underlying theory.

Second, the effect does not, of itself, introduce any explicit chiral symmetry breaking.  Rather, it correlates chiral asymmetries in the Fermi field with those in the propagation of gravitational radiation.

Third, while the effect here has something of the flavor of 
what has been called ``cosmological birefringence,'' and the related suggestion of Lue, Wang and Kamionkowski that gravitational waves of differing circular polarizations might propagate differently \cite{LWK}, there are substantial differences (beside the fact that those effects involve nonstandard terms in the Lagrangian):  ``cosmological birefringence'' would be an effect on electromagnetic, not gravitational, waves; the suggestion of Lue et al. did concern gravitational waves, but gave rise to a circular dichroism (unequal propagation of the polarizations' amplitudes) rather than a gyrotropy.

I should point out at this juncture that the gyrotropy (\ref{phrot}) is not due to a circular birefringence, that is, a difference in the speeds of propagation of the polarizations.  Each polarization propagates with the speed of light.  The effect is in this sense subtler than birefringence.
Mathematically, the characteristics of the waves (the wave fronts) remain null hypersurfaces, and so move at the speed of light; but the transport equations for the wave profiles along the bicharacteristic curves (that is, along the null geodesics ruling those fronts) are sensitive to the polarization, so the form of the profile as it advances along the front changes.\footnote{So, while circular birefringence can cause gyrotropy, here the cause is different.
Not all workers distinguish between gyrotropy and circular birefringence;  indeed, what is
usually called ``cosmomological birefringence'' would be called a gyrotropy of electromagnetic waves here.}

Finally, the link between chirality and space--time curvature calls to mind the gravitational anomaly 
\begin{equation}\label{DSeek}
\nabla ^aj_a^5 =\frac{1}{192\pi ^2}R_{abcd} \, {}^*R^{abcd}
\end{equation}
first derived by Delbourgo and Salam \cite{DelbourgoSalam}.  
This is nonlinear in the gravitational field, whereas the gyrotropy will be present at the linear level.  And the anomaly is a quantum effect, whereas the gyrotropy is predicted even classically.
But also
the chiral current plays very different roles in the two effects.  Indeed, in eqs. (\ref{phrot}), (\ref{ccd}), the current (times $2\pi\ell _{\rm Pl}^2$) appears to function as a $U(1)$ gauge potential (relative to which the circular polarizations of the gravitational wave have opposite charges) to be integrated along a path, a one-form rather than a three-form:  a dualization, incorporating the Planck area, has entered.  Curiously, there is another case in which a certain chiral current appears mathematically as a one-form:  the chiral isospin current is the gradient of the pion field (times its decay constant) in the PCAC model.

\subsection{Mechanism}

Some features of the gyrotropy require detailed computations, but the general way it arises is not hard to understand.  Space--time physics is described by a coupled system of equations:  Einstein's field equation $G_{ab}=-8\pi G T_{ab}$ (with $G_{ab}$ the Einstein tensor and $T_{ab}$ the stress--energy [including any cosmological constant], with the speed of light taken as unity) together with the evolution equations for the matter present.  We will be concerned with a ``background,'' slowly-varying, configuration on which high-frequency waves, treated as first-order perturbations, may propagate. 

In these circumstances, as is well-known, the perturbation of $G_{ab}$ gives a second-order wave operator on the metric perturbations.  What is special about the stress--energies of fermions is that they involve first derivatives of the metric. (Most stress--energies involve the metric only algebraically.)  This means that the wave equation acquires a non-trivial first-order term from the perturbation of $T_{ab}$.  Because that order is less than the highest-order term, this does not affect the characteristics of the system, and hence the speed of the waves.  But because the order is just one less, it effects the transport equations, that is, the evolution of the wave profiles along the wave fronts.

To see that this affect is proportional to the chiral current requires a calculation based on the explicit form of the fermion stress--energy.  The relevant component, the spin-two part, turns out to be $i{\overline\sigma}'l^aj^5_a$, where the spin-coefficient ${\overline\sigma}'$ is essentially the derivative of the spin-two part of the metric and $l^a$ is the null vector field ruling the wave-fronts.  
The high-frequency condition is used again, and deeply, in this part of the argument.

\subsection{Scale of the gyrotropy today}

Although in principle the gyrotropy should occur very generally, the Planck-area prefactor in eq. (\ref{phrot}) makes its magnitude unobservably small in virtually all circumstances considered in contemporary astrophysical models.  
For instance, even for a supernova emitting $\sim 10^{54}$ neutrinos from a $\sim 10\, {\rm km}$ radius, the chiral column density is $\sim 10^{41}\, {\rm cm}^{-2}$, some twenty-five orders of magnitude below the Planck scale.
Similarly, even if all the dark matter in the Universe were in $\sim 10\, {\rm keV}$ sterile neutrinos, the corresponding number density would be $\sim 10^2\, {\rm cm}^{-3}$, so that the present induced mean cosmic rotation rate would be $\sim 10^{-37}\, {\rm rad}/{\rm Gyr}$.  (Of course, dark matter is not uniformly distributed, but it is very hard to see how a density enhancement could be plausibly large enough to compensate for the smallness of this figure.)  
If black holes do radiate as Hawking \cite{Hawking1974,Hawking1975} suggested, their final bursts could plausibly produce gyrotropies of order unity, but each hole's cross-section would only be of order the Planck area.

\subsection{The very early Universe}

There appears however to be no {\em fundamental} limitation that the gyrotropy should always be negligible.  While the prefactor $\ell _{\rm Pl}^2$ is quite small, one must multiply it by the chiral column density $N_5$ to get the rotation, and that brings in an integration over a null geodesic which may extend even over cosmological scales.  The smallness of the effects in the previous paragraph is  due to the difficulty in accumulating significant chiral column densities by known means in the current Universe; it is not an absolute restriction.

The remarkable simplicity and naturality of the formula (\ref{phrot}) is a strong reason for thinking that this effect may well be consequential for physics:  it is plausibly a hint that there is important chiral physics involving gravity in the very early Universe.  The energy scales would have to be high enough that a significant chiral column density was present, and there would have to be gravitational radiation.
While this differs from most contemporary modeling, it is not so far removed from some ideas, for instance the 
``gravitational leptogenesis'' scenario proposed by Alexander et al. \cite{APSJ}.
These authors introduce a chiral symmetry breaking by coupling a complex inflaton field to the curvature.  This creates  elliptically polarized gravitational waves, which are then sources for a chiral current, as in eq. (\ref{DSeek}).  A chiral density of leptons is thus generated very early.

Here I will focus, not on modeling the very early Universe, but on general features of the gyrotropy, and on the possibility that it might have left imprints on physics which would be detectable in the near future.
I will consider 
the possibility that there was at some period in the very early Universe a significant chiral number density through which primordial gravitational waves passed, leaving a polarization effect which might be detectable in the near future.
Because $j^5_a$ is parity-odd, the gyrotropies will interconvert E- and B-polarizations of gravitational waves, and a characteristic prediction of this effect would be the creation of E-B correlations from E-E or B-B ones.  There would also be, in general (for spatially inhomogeneous chiral densities), a creation of circularly polarized gravitational waves.

There will be consequences for the cosmic microwave background (CMB), because of the correlations in the gravitational waves passing through it (and not because of any other new physics).  There will be no creation of circular CMB polarization, because there is no mechanism for that in the standard model.  On the other hand, there will be 
a creation of E-B and T-B correlations in the CMB.  (See ref. \cite{KamionkowskiKovetz,CNSTZ} for current prospects for the detection of cosmic microwave background B-modes.)
That we get such correlations is not really a surprise (since any change linear in $j^5_a$ must be parity-odd), but the details of the mechanism point to a more primitive issue.

\subsection{Violating cosmic coherence}

In most conventional models, there are important restrictions
on the subspace of admissible data for primordial gravitational waves.  These restrictions appear most explicitly in inflationary arguments, where the waves are supposed to be squeezed into a particular subspace, which then serves as an initial-data space for waves in a hot big bang.  Albrecht \cite{Albrecht2001} called this restriction ``cosmic coherence.''  
I will call waves from this squeezed subspace ``beta'' modes, and the complementary ones ``alpha'' modes.  
(While one might have interchanged these terms on the grounds that the beta modes are the more familiar ones, the usage here is a little more natural from the point of view of the concepts involved.
Alpha modes would 
generically be present, and their vanishing is precisely the reason that that the gravitational-radiation entropy of the initial state is very low.)

The restriction to beta modes amounts to a phase coherence of the left- and right-circularly polarized gravitational waves.
Gravitational gyrotropy breaks this coherence.  That is, initially coherent waves, propagating through a chiral density, will decohere.  Put another way, if we examine the waves at this later point, and imagine running them back in time through a Universe {\em without} a chiral density, they would appear not to have come from the squeezed subspace, that is, they would contain alpha-modes.
This difference would in principle be detectable in the present day if the gravitational waves could be measured.  It also will affect the CMB, since it means a different set of gravitational waves will propagate through the matter and radiation, and contribute different anisotropies.  

I work out the consequences for the CMB correlations $C^{XY}_l$ (where $X$, $Y$ are E, B, T, and $l$ is the multipole index) in the case of a spatially
uniform chiral current in the cosmic time direction in a $k=0$ Friedmann--Robinson--Walker walker space--time, accumulating before the usual hot big bang.  In this case, only one new parameter, the chiral column density $N_5$, enters.  
A simple formula (\ref{newcorr}) is found for the correlations, periodic in $N_5$ with period $2\pi$ in $\theta =8\pi \ell _{\rm Pl}^2 N_5$, cycling through a response purely to beta modes (that is, the usual formulas) at even multiples of $\pi$ and purely to alpha modes at odd multiples.\footnote{The quadrupling of the period relative to eq. (\ref{phrot}) comes from two factors or two, the first being the spin-two character of the field and the second the quadratic form of the correlation functions.}  

Several interesting classes of potential effects are contained here:
(a) those which are qualitatively new, the correlations $C^{\rm TB}_l\not= 0$, $C^{\rm EB}_l\not=0$, which would be forbidden by parity in standard models;
(b) the contributions of the alpha modes even to the standard, matched-parity, correlations, although these are at least second-order in $\ell _{\rm Pl}^2N_5$;
(c)~the potential of the alpha-mode contributions to detect physics at earlier times (in particular, as we shall see, deeper within the tight-coupling era) than do the beta-modes, since the alpha-modes tend to dominate at earlier times.

Because alpha-modes dominate beta-modes at sufficiently early times (by a factor 
$\sim (k\eta )^{-1}$, with $k$ the comoving wavenumber at $\eta$ the conformal time), the alpha-mode effects on the CMB could potentially be significantly larger than beta-mode ones, particularly at low wave-numbers.  On the other hand, the gravitational waves do not have much effect until the the opacity of the plasma drops.  So we might expect enhanced effects at wave-numbers below about the inverse conformal time of recombination, and multipoles $l\lesssim 50 -100$ --- weighted by $\sin \theta$, of course.
This turns out to be roughly correct.

To estimate the sizes of the effects, some numerical investigations were made with the CLASS program~\cite{CLASSII,LegourgesTram2013}, modified to compute the alpha-mode transfer functions.  
The rough picture sketched above is confirmed.  

A detailed statistical analysis to constrain $\theta$ would be a substantial project, and was not undertaken.  Instead, the fact that the alpha-mode effects were most distinctive at very low multipoles was used to get some crude estimates.  The effects are then proportional to the amplitude of the scalar perturbations, the tensor-to-scalar ratio $r$, and $\sin\theta$ or $\sin ^2\theta /2$.  If $r\gtrsim .1$, then data from Planck~\cite{Planck}, make values of $|\theta |$ larger than around $.4$ (modulo $2\pi$) unlikely, but if $r\lesssim .01$, no significant constraints were found.  More refined analyses may provide more information.

I briefly discuss one more consequence in principle of the gyrotropy.  This is that it provides a mechanism for the generation B-mode Bondi shear whenever there are memory effects; this B-mode shear can be thought of as the specific (per unit mass) spin angular momentum of gravitational radiation emitted from a system \cite{ADH2007}.

\subsection{Organization}

Evidently special properties of Fermi fields underlie all the effects discussed here, and evidently what must enter (since the fields couple to gravity via Einstein's equation) is the fermionic stress--energy, that is, certain tensorial formulas derived from spinorial expressions.  However, once these results are in hand the spinors are not explicitly needed.
Because of this, and because such derivations are rather technical and specialized, all of the explicitly spinorial computations are placed in an appendix,
which can be omitted by the reader willing to simply accept its results.  For readers interested in the technicalities, it is worth noting that the computation is done by two-spinor techniques, which are here more efficient than four-spinors and gamma matrices.  The relation between the approaches is given explicitly.  The appendix can be read after section 2.2.

Section II explains the mechanism driving the rotation (making use of the results of the appendix).  
No explicitly cosmological assumptions are made at this stage; the treatment is general.  This generality is not only desirable of itself, but also makes clear the nature of the high-frequency expansion which is used, and what its limitations are.  However, the details of the computations in this section are not used subsequently, so the reader interested primarily in the results can safely skim this.  
The main formulas needed later are:  the definition of the expansion (in the beginning of subsection B, up to eq. (\ref{metexp})); the expression for the lead term, eqs. (\ref{hpert}), (\ref{firstmdef}) and the accompanying paragraph, and eqs. (\ref{tranintphi}), (\ref{tranintpsi}) showing the appearance of the chiral current in the transport integrals.

Section III gives the general set-up for the application of the results to a 
$k=0$  Friedmann--Robertson--Walker
cosmological model, where primordial gravitational waves encounter a chiral density in the very early Universe. 

Section IV works out the case of a spatially uniform chiral density.  Formulas for the CMB correlations $C^{XY}_l$ are derived; these new effects enter only through a single parameter $\theta =8\pi \ell _{\rm Pl}^2N_5$, and the results are $2\pi$-periodic in it.  Some numerical investigations are made with a slightly modified CLASS
program.
The comparison with data from Planck is made, and very rough estimates of constraints on $\theta$ given.
Also here the sensitivity of alpha-mode effects to the tight-coupling regime is uncovered.

In Section V, inhomogeneous chiral densities are considered.  These are technically more difficult to treat, and the analysis is only done to first order in $j_5^a$.  A main effect of inhomogeneities is to smear out the correlations; this is because the gravitational waves scatter off the inhomogeneities.  Because $j_5^a$ is parity-odd, the only non-trivial first-order effects on $C^{XY}_\ell$ can be for $X$ and $Y$ of opposite parities.  However, because of the invariance of the transfer functions, the CMB correlations can detect only the part of $N_5(\bx )$ whose spatial dependence is even.
On the other hand, in this section it is shown that an inhomogeneous chiral density will in general induce correlations in circular polarization on the gravitational waves, and these are sensitive to the part of $N_5$ with odd spatial dependence.

Section VI contains a discussion of the results.

There are two appendixes.  Appendix A, as already noted, derives the perturbation of the fermion stress--energy needed for the analysis.  Appendix B briefly indicates how the effects here could be treated in nonlinear gravity without invoking a background-plus-perturbation approach, and so gives a more fully relativistic perspective.

{\em Conventions.}
The metric has signature $+{}-{}-{}-$; tensorial indices are small Latin letters; the indices $j$, $k$ are reserved for three-tensors.  The curvatures obey $[\nabla _a,\nabla _b] v^d =R_{abc}{}^dv^c$, $R_{ac}=R_{abc}{}^b$, $R=R_a{}^a$ and Einstein's equation is $G_{ac}=R_{ac}-(1/2) Rg_{ac} =-8\pi GT_{ac}$, with $G$ being Newton's constant, the stress--energy being $T_{ac}$ (and any cosmological constant is absorbed in $T_{ac}$).  

The tensorial conventions all accord with those in Penrose and Rindler \cite{PR1984,PR1986}.  Spinors only appear explicitly in the Appendixes, and conventions for them are discussed there.

The symbol $k$ is used in a few places for the the curvature indicator for Friedmann--Robertson--Walker space--times (here only $k=0$ is considered), but usually it stands for 
the magnitude of the comoving wave-number.  Context should make the meaning clear.

\section{Mechanism}

In this section the rotation of the polarization of gravitational waves will be derived.  
While some of the calculations are lengthy and technical, the main idea is straightforward.

While in the most general circumstances there is no simple way to isolate the wavelike degrees of freedom of a gravitational system from others, in many situations it is legitimate to model the waves as (relatively) rapidly changing linear perturbations on a given background.  If we consider a system with matter as well as gravity, it will be described by a set of coupled equations:
\begin{equation}
\left.\begin{aligned}
  &G_{ac}=-8\pi G T_{ac}\\
  &\text{equations of motion for the matter}
  \end{aligned}\right\}\, ,
\end{equation}
where $G_{ac}$ is the Einstein tensor and $T_{ac}$ is the stress--energy (including any cosmological constant term).
The first-order perturbations of this will have the form 
\begin{equation}\label{peom}
\left.\begin{aligned}
  &\delta G_{ac}=-8\pi G\delta T_{ac}\\
  &\text{first-order perturbed matter equations of motion}
  \end{aligned}\right\}\, .
\end{equation}
In general these are coupled.
Here $\delta G_{ac}$ is a second-order wave operator acting on the metric perturbation $\delta g_{ac}$.  The quantity $\delta T_{ac}$ in general contains two sorts of terms:  there are those 
$\delta _{\rm matter}T_{ac}$ coming from perturbations of the matter fields figuring in $T_{ac}$, and there are those $\delta _g T_{ac}$  coming from the dependence of $T_{ac}$ on the metric.  It is these second sorts of terms which are responsible for the novel effects.

What is special about fermions is that their stress--energies involve first derivatives of the metric.  (By contrast, minimally coupled scalar fields', and electromagnetic and Yang--Mills fields', stress--energies only depend on the metric algebraically.)    This means that the right-hand side of the top line of (\ref{peom}) contributes first-derivative terms to the wave equation for $\delta g_{ac}$.  Since the highest- (second-) order terms are unaffected, the characteristics of the system are null hypersurfaces, as usual, and the wave profiles can be taken (exploiting the gauge freedom) to be transverse in the usual sense.  Since the characteristics are null, the waves propagate at the speed of light.  However, the transport equations, which govern the propagation of those profiles along the null geodesics ruling the characteristics, {\em are} affected by the first-order terms \cite{CourantHilbertII}.  In other words, even though the waves' speed is unaffected, what is sometimes called the Huygens term, that is, the form of the wave-front, is changed.  The coefficient of the first-order term, which determines this change, involves the chiral current $j^5_a$.

I have not so far restricted the matter terms, and one could imagine different first-order perturbations of the matter (compatible with the high-frequency character of the metric perturbations).  However, I assume further that the matter is unperturbed except for its response to the gravitational waves.  This assumption is implemented in the analysis in Appendix A.

The distinction between gyrotropy and birefringence can be made at this point.  Birefringence generally refers to different indices of refraction for different polarizations; mathematically, this corresponds to polarization-dependent highest-order terms in the propagation equation, which govern the speed of the wave.  A circular birefringence thus will evidently give rise to a rotation of polarization, that is, a gyrotropy.  However, we see that the gyrotropy here does not come about for this reason, because the highest-order terms are unchanged.  Gyrotropy without birefringence is possible because the waves are transverse.

\subsection{First-order perturbations of geometry}

Now consider linear perturbations of an arbitrary ``background'' space--time with metric $g_{ab}$.  If this is perturbed by $\delta g_{ab}=h_{ab}$, the change in the connection is
\begin{equation}
\delta \Gamma _{bc}{}^d =\frac{1}{2}\left[
  \nabla _b h_c{}^d+\nabla _ch_b{}^d-\nabla ^dh_{bc}\right]\, .
\end{equation} 
This gives a change
\begin{eqnarray}
\delta R_{ac}&=&R_{p(a}h_{c)}{}^p -R_{(a|b|c)}{}^qh_q{}^b\\
&& +\frac{1}{2}\left( \nabla _a\nabla _ch -2\nabla _{(c}\nabla _{|b|}h_{a)}{}^b +\nabla ^2h_{ac}\right)\, ,\nonumber
\end{eqnarray}
in the Ricci tensor and
\begin{eqnarray}
\delta G_{ac}&=&R_{p(a}h_{c)}{}^p -R_{(a|b|c)q}h^{qb}\\ &&
+\frac{1}{2}\left( \nabla _a\nabla _c h -2\nabla _{(c}\nabla _{|b|} h_{a)}{}^b +\nabla ^2h_{ac}\right)\nonumber\\
&&-\frac{1}{2} h_{ac}R +\frac{1}{2}g_{ac} \left( R_{pq}h^{pq}
  -\nabla ^2h +\nabla ^p\nabla ^qh_{pq}\right)\nonumber
\end{eqnarray}   
in the Einstein tensor, where $h=h_a{}^a$.
(Here vertical strokes set off postscripts not included in the symmetrizations.)

\subsection{High-frequency expansion}

The high-frequency expansion will be done in an invariant way, although it will depend on a choice of a foliating family of smooth null hypersurfaces, which will be the wave-fronts.  Let these hypersurfaces be the level sets of a function $u$ (increasing towards the future, and having dimension time).  
Then $l_a=\nabla _au$ is the null normal, and one has $l^b\nabla _bl^a=0$, so $l^a$ is the parallel-propagated null tangent to the null geodesics ruling the hypersurfaces.  

It is convenient to complete $l^a$ to a normalized, oriented, null tetrad $l^a$, $m^a$, ${\overline m}^a$, $n^a$, where
\begin{equation}
  l^an_a=-m^a{\overline m}_a =1
\end{equation}
(and all other inner products among them vanish).  The orientation is such that, at any point we could choose a standard basis with 
\begin{eqnarray}
  l^a&=&2^{-1/2}\left(\partial _t+\partial _z\right)\label{basel}\\
  n^a&=&2^{-1/2}\left(\partial _t-\partial _z\right)\label{basem}\\
  m^a&=&2^{-1/2}\left( \partial _x -i\partial _y\right)\label{basen}\, .
\end{eqnarray}
We may choose this basis to be transported parallel along $l^a$.
(It is not quite uniquely determined; one could perform a null rotation about $l^a$.  Note that the $x$, $y$, $z$ figuring here could not generally be comoving coordinates in a cosmological application.)

Let us put
\begin{equation}\label{metexp}
\delta g_{ab}  \sim e^{-i\omega u}\left(
  h^{(0)}_{ab}+\frac{h^{(1)}_{ab}}{-i\omega} 
+\frac{h^{(2)}_{ab}}{(-i\omega )^2} 
+\cdots\right)
   +\text{conjugate}\, ;
\end{equation}
we are interested in this in the limit of large $\omega$.    
Notice that a covector field
\begin{equation}
\xi _a \sim e^{-i\omega u}\left( \xi ^{(0)}_a +\cdots\right)
  +\text{conjugate}\, ,
\end{equation}
would give a  gauge transformation
\begin{eqnarray}
2\nabla _{(a}\xi _{b)}&\sim&
2e^{-i\omega u}\left( -i\omega l_{(a}\xi ^{(0)}_{b)}
  +\left(\nabla _{(a}\xi ^{(0)}_{b)} +l_{(a}\xi ^{(1)}_{b)}
  \right)
  +\cdots\right)\nonumber\\
  &&+\text{conjugate}\, ,
\end{eqnarray}     
and thus by taking $\xi ^{(0)}_a=0$ and adjusting $\xi ^{(1)}_a$, we may eliminate any term in $h_{ab}^{(0)}$ proportional to a symmetrization of $l_a$ with another covector.

Now one expands the Einstein tensor and the stress--energy in inverse powers of $\omega$.  The perturbation of the Einstein tensor is
\begin{eqnarray}\label{EEk}
\delta G_{ab}&=&
e^{-i\omega u}\left( G^{(-2)}_{ab}(-i\omega )^2 +G^{(-1)}_{ab} (-i\omega )+\cdots\right)\nonumber\\
&& +\text{conjugate}\, .
\end{eqnarray}
The stress--energy is treated in the appendix.

\subsubsection{The order $(-i\omega )^2$ term}

As noted above, the order $(-i\omega )^2$ term for the perturbed stress--energy vanishes, and then the leading term in eq. (\ref{EEk}) gives an algebraic restriction on $h^{(0)}_{ab}$.  Precisely, the term
\begin{equation}\label{Gmtwo}
 G_{ac}^{(-2)} =
  \frac{1}{2}\left( l_al_c h^{(0)}-2l_{(a}l_{|b|} h^{(0)}_{c)}{}^b
   +g_{ac}(l^pl^qh^{(0)}_{pq})\right)
\end{equation}
must vanish, and it is an algebraic exercise to verify this forces $h^{(0)}_{ab}$ to be transverse and trace-free (modulo gauge).  We may therefore discard gauge terms and write
\begin{equation}\label{hpert}
  h^{(0)}_{ab}=\phi m_am_b +\psi {\overline m}_a{\overline m}_b\, ,
\end{equation}
for some functions $\phi$, $\psi$.  (Here $\phi$ and $\psi$ are not scalar metric perturbations, but the spin-weighted coefficients of tensor perturbations.)

Here note that, in terms of the basis (\ref{basel})--(\ref{basen}) (which, recall, is not comoving), we have
\begin{equation}\label{firstmdef}
  m_am_b=\frac{1}{2}\left[ (dxdx-dydy) -i(dxdy+dydx)\right]\, 
\end{equation}  
expressed the polarization in terms of the usual $e_+=(dxdx-dydy)$ and $e_\times =(dxdy+dydx)$.  (This is the \emph{left} circular polarization tensor in the sense of Misner, Thorne and Wheeler~\cite{MTW}, but would be the gravitational analog of \emph{right} circular optical polarization in the sense of Jackson~\cite{Jackson1967}.  The contribution of the $e^{-i\omega u}\phi m_am_b$ term to the Weyl tensor is left-handed in the usual particle-physics sense, and that of $e^{-i\omega u}\psi {\overline m}_a{\overline m}_b$ is right-handed.)

\subsubsection{The order $(-i\omega )$ term}

The next, order $(-i\omega )$, term in the expansion is more complicated.  The equation
\begin{equation}\label{feek}
  G^{(-1)}_{ac}=-8\pi G T^{(-1)}_{ac}
\end{equation}
is a system of ten scalar equations involving $h^{(0)}_{ac}$ and $h^{(1)}_{ac}$.  I shall in this section only consider the left-hand side.

Explicitly, the term is
\begin{widetext}
\begin{eqnarray}\label{Gmone}
  G^{(-1)}_{ac}&=&\frac{1}{2}\left( \nabla _a(l_ch^{(0)})+l_a\nabla _ch^{(0)}
    -2(\nabla _{(a}l_{|b|}h^{(0)}_{c)}{}^b 
+l_{(a}\nabla_{|b|}h^{(0)}_{c)}{}^b)
+\nabla ^p(l_ph^{(0)}_{ac})
    +l_p\nabla ^ph^{(0)}_{ac}\right)\nonumber\\
    &&+\frac{1}{2}g_{ac}\left( -\nabla _p(l^ph^{(0)})-l_p\nabla ^ph^{(0)}
      +\nabla ^p(l^qh^{(0)}_{pq}) +l^p\nabla ^qh^{(0)}_{pq}\right)
      +\frac{1}{2}\left( l_al_ch^{(1)} -2l_{(c}l_{|b|}h^{(1)}_{c)}{}^b +g_{ac}l^pl^qh^{(1)}_{pq}\right)\, .\qquad
\end{eqnarray}  
\end{widetext}
Note that the last line here has the same form as eq. (\ref{Gmtwo}), but with $h^{(1)}_{ab}$ replacing $h^{(0)}_{ab}$.
This means that, if $T^{(-1)}_{ab}$ were known, one could solve
certain components of eq. (\ref{feek}) algebraically for $h^{(1)}_{ab}$ (modulo gauge) in terms of $h^{(0)}_{ab}$ and $T^{(-1)}_{ab}$.
In fact, it is not hard to check, given the form (\ref{hpert}) of $h^{(0)}_{ab}$, that the only terms in (\ref{Gmone}) that cannot be influenced by the choice of $h^{(1)}_{ab}$ are those proportional to $m_am_c$ and ${\overline m}_a{\overline m}_c$.  These will turn out to give us the transport equations.

We have, then, for the left-hand sides of the transport equations,
\begin{eqnarray}
  {\overline m}^a{\overline m}^cG^{(-1)}_{ac}
  &=&l\cdot\nabla\phi +(1/2)(\nabla\cdot l)\phi\\
   {m}^a{m}^cG^{(-1)}_{ac}
  &=&l\cdot\nabla\psi +(1/2)(\nabla\cdot l)\psi\, .
\end{eqnarray}

\subsection{The transport equations}

We may now assemble the results to get the transport equations. 
These are the $m^am^a$ and ${\overline m}^a{\overline m}^a$ components of the next-to-leading order term in the high-frequency expansion of the Einstein equation $G^{(-1)}_{ac}=-8\pi GT^{(-1)}_{ac}$.  These components of the Einstein tensor were derived in the previous subsection; of the the stress--energy, in the appendix.  

The transport equations are:
\begin{eqnarray}
l\cdot\nabla\phi +(1/2)(\nabla\cdot l)\phi
  &=&-4\pi G il\cdot j^5 \phi\\
 l\cdot\nabla\psi +(1/2)(\nabla\cdot l)\psi
  &=&+4\pi G il\cdot j^5 \psi\, .
\end{eqnarray}  
So the evolution of the perturbations along the geodesic is given by
\begin{eqnarray}
  \phi (\gamma (s)) &=&\phi (\gamma (s_0))\nonumber\\
&&\times    \exp \int _{s_0}^s \left[ -(1/2)\nabla\cdot l -4\pi iGl\cdot j^5\right]\, ds\qquad\label{tranintphi}\\
\psi (\gamma (s)) &=&\psi (\gamma (s_0))\nonumber\\
&&\times    \exp \int _{s_0}^s \left[ -(1/2)\nabla\cdot l +4\pi iGl\cdot j^5\right]\, ds\, .\qquad\label{tranintpsi}
\end{eqnarray}
In these, the real factor $\nabla\cdot l$ gives the usual amplification or attenuation due to the convergence or divergence of the null geodesics.  However, it is the phase factors which are of interest here.      

Thus the gravitational wave would have the coefficients of its polarization tensors $m_am_b$, ${\overline m}_a{\overline m}_b$ gain phases $\mp 4\pi G\int j^5_ad\gamma ^a$ (writing now $l^a ds=d\gamma ^a$), which is the effect a rotation by $-2\pi G\int j^5_ad\gamma ^a$, looking along the axis of propagation into the oncoming wave, would have. 
By its construction, the rotation is measured with respect to parallel transport.

\section{Gravitational gyrotropy in cosmology}

In this section, I adapt the foregoing results to perturbations of a $k=0$ Friedmann--Robertson--Walker space--time.  The goal is to connect with the notations and conventions common in cosmology; computations for particular classes of chiral distributions will be done in subsequent sections.

However, even this section will have to go a bit beyond the current conventional treatment.  This is because, as noted in the introduction, gravitational gyrotropy will generally violate the `cosmic coherence' assumption which is usually made.
This assumption --- which is generally motivated by inflation --- requires that the waves come from a particular squeezed subspace at the beginning of the hot big bang, a subspace with phase coherence between the right- and left-circularly polarized modes.  
If, some time after the big bang but still in the very early Universe, such gravitational waves pass through a chiral column density, the initial phase coherence will be lost.  This would mean that these later waves, if propagated backwards in time through a space--time {\em not} containing the chiral column density, would appear to have come from a different subspace. 
In fact, phrasing the analysis in terms of such altered initial data is the most convenient thing to do, so we need terminology and conventions suitable for this.  The usual modes will be called beta modes; their natural complement will be the alpha modes.

\subsection{Preliminaries}

I will now use comoving coordinates $x^j$, and a positive-definite standard Euclidean three-metric $\varepsilon _{jk}$ used to raise and lower Euclidean indices; for most purposes I use a three-vector notation.  Usually a conformal time $\eta$ is used, although in some places physical time $t$ is more natural.  The metric is then
\begin{eqnarray}
  ds^2&=&a(\eta )^2\left( d\eta ^2-\varepsilon _{jk}dx^jdx^k\right)\\
&=&dt ^2-a^2\varepsilon _{jk}dx^jdx^k\, .
\end{eqnarray}
For the perturbations, we will take the circular frequency $\omega=k\geq 0$ and the phase function
\begin{equation}
  ku=k(\eta -{\hat\bk}\cdot\bx )\, ,
\end{equation}
where $\bk =k\hat\bk$ is the spatial wave-vector (hats indicate comoving unit vectors).
Then the tangent covector is $l_a=\nabla _au =d\eta -{\hat\bk}\cdot d\bx$ and $l^a=a^{-2}\left( \partial _\eta +{\hat\bk}\cdot{\bm\nabla}\right)$.

For the polarizations, 
one has
\begin{equation}\label{polter}
  m_am_b =\frac{1}{2} a^2 (u_j u_k -v_j v_k -i(u_jv_k+v_ju_k)) dx^j dx^k
\end{equation}
for suitable constant unit vectors $\bm u$, $\bm v$ (with $\bm u$, $\bm v$, $\hat\bk$ forming a Euclidean right-handed comoving-orthonormal triad).  Thus we could write
\begin{equation}
  m_am_b =\frac{1}{2}a^2 (\epsilon ^{+}_{ab}-i\epsilon ^{\times}_{ab})\, ,
\end{equation}
in terms of the plus and cross polarizations adapted to the comoving frame.

There is a slight subtlety here, which is the dependence of the polarization tensors on the vector $\bk$ and its consequent behavior under the inversion $\bk\to -\bk$, which will be important shortly.  On one hand, there must be a dependence (for the polarization is orthogonal to $\bk$); on the other, there is, for each $\bk$, the freedom to make a phase rotation in $m_a$.  There is no uniform choice in the literature for this.  I follow Hu and White \cite{HuWhite1997}, who choose ${\bm u}$ proportional to $\partial _\theta$ and $\bm v$ proportional to $-\partial _\phi$ on the sphere, which gives
\begin{equation}
  m_am_b(-\bk )= {\overline m}_a{\overline m}_b (\bk )\, ,
\end{equation}
and makes $\epsilon ^{+}_{ab}(\bk )$ parity-even and $\epsilon ^{\times}_{ab}(\bk )$ parity-odd.  
It should be emphasized that this is simply a matter of convention, and that one could equally well include a minus sign (or direction-dependent phase factor).  What will matter in the end is, of course, the products of the polarization tensors with the mode functions.  However, just for this reason, this convention does very much affect the parities of those mode functions.

The leading,
order-$(-i\omega )^0$, term in the metric perturbation is
\begin{eqnarray}\label{pertform}
\delta g_{ab}&\sim& \int d^3\bk\, e^{-i(k\eta -\bk\cdot\bx)}
\left( \phi (\bk ,\eta )m_am_b+\psi (\bk ,\eta ){\overline m}_a{\overline m}_b
\right)\nonumber\\
  &&+\,\text{conjugate}\nonumber\\
&=&\int d^3\bk\, e^{i\bk\cdot\bx}\times\nonumber\\
&&\left[ 
\left( e^{-ik\eta}\phi (\bk ,\eta )+e^{ik\eta}\overline\phi (-\bk ,\eta)\right) m_am_b(\bk )\right.\nonumber\\
&&\left.+\left(e^{-ik\eta}\psi (\bk ,\eta )+e^{ik\eta}\overline\psi (-\bk , \eta)\right){\overline m}_a{\overline m}_b(\bk )
\right]
\, ,
\end{eqnarray}
where the terms in square brackets give the spatial Fourier transform, as usual. 

Now let us turn to
the transport equations (\ref{tranintphi}), (\ref{tranintpsi}).  We have $\nabla _al^a =2a'/a^3$ (with the prime denoting differentiation with respect to $\eta$) but $ds=a^2d\eta$, and so
\begin{equation}
\exp -\frac{1}{2}\int \nabla _al^a\, ds =a^{-1}
\end{equation}
for the transport factor accounting for the divergence of the congruence.

\subsection{Primordial waves}

It is usually assumed that a stochastic background of gravitational-wave perturbations, in a particular Lagrange subspace, accompanies the big bang.  
As noted earlier, however, waves which subsequently pass through a chiral density will be affected by a gyrotropy, and this makes them appear, to later observers not accounting for the gyrotropy, as having come from a different subspace.  This point of view will be useful in making contact with CMB modeling.  We therefore need to consider primordial waves which may not come from the usual squeezed subspace.

To see how to proceed, it will be helpful to connect with some of the standard notation figuring in treatments of the CMB.
Mode functions $H^{(\pm 2)}$ compatible with the notation of Hu and White \cite{HuWhite1997} are
\begin{eqnarray}
H^{(2)}(\bk ,\eta ) &=&\sqrt{2/3}\left( e^{-ik\eta}\psi (\bk ,\eta )+e^{ik\eta}{\overline\psi}(-\bk ,\eta )\right)\, ,\qquad\label{Hplus}\\
H^{(-2)}(\bk ,\eta ) &=&\sqrt{2/3}\left( e^{-ik\eta}\phi (\bk ,\eta )+e^{ik\eta}{\overline\phi}(-\bk ,\eta )\right)\, .\qquad\label{Hminus}
\end{eqnarray}
These are to be understood as stochastic variables.  (Hu and White did not work explicitly with stochastic variables, instead using mode functions of the scalars $(k,\eta )$ and probability distributions; the explicit representation is due to Seljak and Zaldarriaga \cite{ZS1997}, although not in quite this notation.)  

In inflationary theory, during the inflationary period the modes freeze (and are not wavelike gravitational excitations), and then begin to change again after inflation ends.  This condition may be expressed as
\begin{equation}\label{reeq}
\partial _\eta H^{(\pm 2)}(\bk ,\eta )\Bigr|_{\eta=\eta_{\rm e}}=0\, ,
\end{equation} 
where $\eta _{\rm e}$ 
is the conformal time at which inflation ends.  This is a very strong constraint on the data for the waves (and not just on the probability distribution).  It restricts them to a particular Lagrange subspace.  
When inflation ends, some sort of pre-heating and reheating is supposed to occur, giving a transition to a radiation-dominated expansion, at, say $\eta _{\rm rad}$.  In the transition period $\eta _{\rm e}\leq\eta\leq\eta _{\rm rad}$, in principle, the gravitational waves could be significantly affected by the precise nature of the space--time through which they pass:  there is some evolution taking the data $(H^{(\pm 2)}(\bk ,\eta _{\rm e}),\partial _\eta H^{(\pm 2)}(\bk ,\eta _{\rm e}))$ at $\eta _{\rm e}$ to those at $\eta _{\rm rad}$, which depends on the physics of the transition.

In practice, it is usually assumed that the data at $\eta _{\rm rad}$ can be set by a limiting case of eq. (\ref{reeq}).  That is, one works with a radiation-dominated model beginning at conformal time $\eta _{\rm rad}=0$, and takes the initial data for the gravitational waves to be constrained by 
\begin{equation}\label{efreeq}
  \lim _{\eta\downarrow 0} \partial _\eta H^{(\pm 2)}(\bk ,\eta ) =0\, .
\end{equation}  
This assumption is used very strongly in contemporary CMB modeling, and is the justification for working with transfer functions giving the response of temperature and polarization to the (limiting) initial data for the value of the tensor metric perturbation and ignoring the transfer functions for the response to the (limiting) initial value for its time-derivative.

We have in mind a situation where, sometime after the big bang but still in the very early Universe (in particular, in the radiation era), primordial gravitational waves passed through a chiral current.  While the physical waves themselves will be supposed to obey the constraint, their appearance after passing through the chiral current, will be the same as waves not having passed through such a current but also not obeying the constraint.  We thus wish to consider data not obeying the constraint, even though the solution those data correspond to is the physically correct one only after the encounter with the chiral current.

In the radiation-dominated era, the high-frequency expansion is a good one.  Taking into account the computation of the transport effects at the end of the last subsection, we will have
\begin{equation}\label{phipsiinit}
\left.\begin{aligned}
  \phi (\bk ,\eta )&=\frac{a'(0)}{a(\eta )}\lambda (\bk)\\
  \psi (\bk ,\eta ) &=\frac{a'(0)}{a(\eta )}\rho (\bk)
  \end{aligned}\right\}
   \text{before encountering\ } j_5^a\, ,
\end{equation}
where $\lambda$ and $\rho$, depending only on $\bk$, may be regarded as giving the initial left- and right-circularly polarized densities of gravitational waves, and the factor $a'(0)$ is included to make the relation independent of the overall scale of $a$.

Because the modes generally diverge as $\eta ^{-1}$ as $\eta \downarrow 0$,\footnote{It is worth recalling, however, that to get the metric perturbations one must multiply by the square of the conformal factor, that is, by $\sim \eta ^2$, so all of these waves are bounded in the usual chart.} it is easiest to work with $\eta  H^{(\pm 2)}(\bk ,\eta )$, and identify the initial data
\begin{eqnarray}
  \alpha^{(2)} (\bk ) &=&\lim _{\eta \downarrow 0} \eta  H^{(2)}(\bk ,\eta )\nonumber\\
&=&\sqrt{2/3} (\rho (\bk )+{\overline\rho}(-\bk ))\label{wdatone}\\
\beta ^{(2)} (\bk ) &=&\lim _{\eta \downarrow 0}\partial _\eta \eta  H^{(2)}(\bk ,\eta )\nonumber\\
&=&-ik\sqrt{2/3} (\rho (\bk )-\overline\rho (-\bk ))\\
\alpha^{(-2)} (\bk ) &=&\lim _{\eta \downarrow 0} \eta  H^{(-2)}(\bk ,\eta )\nonumber\\
&=&\sqrt{2/3} (\lambda (\bk )+{\overline\lambda}(-\bk ))\\
\beta ^{(-2)} (\bk ) &=&\lim _{\eta \downarrow 0}\partial _\eta \eta  H^{(-2)}(\bk ,\eta )\nonumber\\
&=&-ik\sqrt{2/3} (\lambda (\bk )-\overline\lambda (-\bk )  )\label{wdatfour}
\, . 
\end{eqnarray} 
Note that knowledge of these functions is equivalent to knowledge of $\lambda$ and $\rho$.
The usual restriction (implication of ``cosmic coherence'') is that $\alpha ^{(\pm 2)}$ should vanish; the data are then taken to be $\beta ^{(\pm 2)}$, and these accord with the limiting values of $H^{(\pm 2)}$ in the case $\alpha ^{(\pm 2)}=0$.  
We assume as usual that the expectations $\langle\beta ^{(\pm 2)}\rangle$ vanish and we have
\begin{eqnarray}
  \langle\beta ^{(\pm 2)}(\bk )\beta ^{(\pm 2)}(\bak )\rangle &=& {\cal P}(k)\delta (\bk +\bak )\label{coreone}\\
  \langle \beta ^{(2)}(\bk )\beta ^{(-2)}(\bak )\rangle &=&0\, .\label{coretwo}
\end{eqnarray}

\section{The spatially uniform case}

We consider here the case of a spatially uniform chiral current.

\subsection{Formulas for the CMB correlations}

Deriving the formulas for the CMB correlations is largely a matter of assembling previous results.

In the radiation era, if no chiral current is present, the gravitational waves are given by the general form (\ref{phipsiinit}).  The initial data are determined by the relations (\ref{wdatone})--(\ref{wdatfour}).  We make the usual hypothesis that the physical waves correspond to data $\alpha ^{(\pm 2)}=0$, $\langle\beta ^{(\pm 2)}\rangle =0$, and eqs. (\ref{coreone}), (\ref{coretwo}). 
We now regard $\lambda$ and $\rho$ as determined by these initial data; then eq. (\ref{phipsiinit}) applies before the waves encounter the chiral current.
  
After passage through the chiral density, we will have additional phase factors:
\begin{equation}\label{acera}
\left.\begin{aligned}
  \phi (\bk ,\eta )&=\frac{a'(0)}{a(\eta )}e^{-4\pi iGN_5}\lambda (\bk)\\
  \psi (\bk ,\eta ) &=\frac{a'(0)}{a(\eta )}e^{+4\pi iGN_5}\rho (\bk)
  \end{aligned}\right\}
  \text{after encountering } j_5^a \, ,
\end{equation}
where $N_5$ is the chiral column density.
These phase rotations 
have the same effect as would changing the initial data to
\begin{eqnarray}\label{chwavedata}
&&\left[\begin{matrix} \alpha ^{(\pm 2)}_5\\ \beta ^{(\pm 2)}_5\end{matrix}\right]
=\\ 
&&\qquad \left[\begin{matrix}
 \cos (4\pi GN_5) & \mp k^{-1}\sin (4\pi GN_5)\\
   \pm k \sin (4\pi GN_5) &\cos (4\pi GN_5)\end{matrix}\right]
   \left[\begin{matrix} 
\alpha ^{(\pm 2)} \\
 \beta ^{(\pm 2)}\end{matrix}\right]
  \, .\nonumber
\end{eqnarray} 
The gravitational waves, then, are given by evolving the initial data (\ref{chwavedata}).  The waves' correlations follow from those of the data.
Using eqs. (\ref{chwavedata}), (\ref{coreone}), (\ref{coretwo}) and a little algebra,
we may express these as
\begin{widetext}
\begin{eqnarray}\label{modefive}
\Big\langle 
\left[\begin{matrix} \alpha ^{(\pm 2)}_5(\bk )\\ \beta ^{(\pm 2)}_5(\bk )\end{matrix}\right]
\left[\begin{matrix} \alpha ^{(\pm 2)}_5(\bak )& \beta ^{(\pm 2)}_5(\bak )\end{matrix}\right]
\Big\rangle &=&
  \left[\begin{matrix} 
     k^{-2}\sin ^2 (4\pi GN_5) &\mp k^{-1} \sin (4\pi GN_5)\cos (4\pi GN_5)\\
     \mp k^{-1} \sin (4\pi GN_5)\cos (4\pi GN_5) &\cos ^2(4\pi GN_5)
      \end{matrix}\right]
      \nonumber\\
      &&\times
       {\mathcal P}(k) \delta (\bk +\bak)\, ;
\end{eqnarray}      
\end{widetext}
there are no correlations between quantities with superscripts of opposite signs.

We now turn to the gravitational waves' effects on the CMB.
To work this out,
we must introduce transfer functions corresponding to the data $\alpha ^{(\pm 2)}$.  Let $\Delta ^X_{\beta^{(\pm 2)}}$ be the usual transfer function of type $X$ (temperature, E- or B-mode polarization), and set $\Delta ^X_{\alpha ^{(\pm 2)}}$ as the corresponding transfer function for the $\alpha ^{(\pm 2)}$ data.  
Then ignoring, for the moment, the chiral density, the
quantity
$X(\bn )$, depending on the direction $\bn$, will be a stochastic variable
\begin{equation}\label{Xdef}
\sum _{\pm}\int d^3\bk \left( \Delta ^X_{\alpha ^{(\pm 2)}}(\bn ,\bk) \alpha ^{(\pm 2)}(\bk ) 
+\Delta ^X_{\beta ^{(\pm 2)}} (\bn ,\bk)\beta ^{(\pm 2)}(\bk ) \right)\, .
\end{equation}
The usual parity relations
\begin{equation}
\Delta ^{\rm E} _{\beta^{(+2)}} =\Delta ^{\rm E} _{\beta^{(-2)}}
 \quad\text{and}\quad
\Delta ^{\rm B} _{\beta^{(+2)}} =-\Delta ^{\rm B} _{\beta^{(-2)}} \, ,
\end{equation}
and likewise for $\alpha$, hold.
In fact, it will be the transfer functions resolved by multipole, that is $\Delta ^X_{l\beta^{(\pm 2)}}(k)$ and $\Delta ^X_{l\alpha ^{(\pm 2)}}(k)$, that enter in the computations.  The analysis for the alpha-modes is completely parallel to that of the beta-modes; only the initial conditions are different.

To take into account the gravitational waves' passage through the chiral density, assuming that this happens before the usual hot-big-bang physics, we need only replace $\alpha ^{(\pm 2)}$, $\beta ^{(\pm 2)}$ in eq. (\ref{Xdef}) by 
$\alpha ^{(\pm 2)}_5$, $\beta ^{(\pm 2)}_5$.   We will therefore have
\begin{widetext}
\begin{eqnarray}\label{newcorr}
  C_l^{XY}&=&(2l+1)^{-1} \sum _{\pm}\int d^3\bk\, {\mathcal P}(k)
  \left\{
   {\overline\Delta}_{l \alpha^{(\pm 2)}}^X(k) 
     {\Delta}_{l \alpha^{(\pm 2)}}^Y(k) 
     k^{-2}\sin ^2(4\pi GN_5)\right.\nonumber\\
&&     \mp 
     {\overline\Delta}_{l \alpha^{(\pm 2)}}^X(k) 
     {\Delta}_{l \beta^{(\pm 2)}}^Y(k) 
     k^{-1}\sin (4\pi GN_5)\cos (4\pi GN_5)\nonumber\\
 &&     \mp 
     {\overline\Delta}_{l \beta^{(\pm 2)}}^X(k)
     {\Delta}_{l \alpha^{(\pm 2)}}^Y(k) 
     k^{-1}\sin (4\pi GN_5)\cos (4\pi GN_5)\nonumber\\
    &&  \left.+
     {\overline\Delta}_{l \beta^{(\pm 2)}}^X(k) 
     {\Delta}_{l \beta^{(\pm 2)}}^Y(k) 
     \cos ^2 (4\pi GN_5)\right\}\nonumber\\
     &=&(2l+1)^{-1} \int d^3\bk \,{\mathcal P}(k)
  \left\{
  (1+\pi _X\pi _Y)
   {\overline\Delta}_{l \alpha^{(+ 2)}}^X(k) 
     {\Delta}_{l \alpha^{(+ 2)}}^Y(k) 
     k^{-2}\sin ^2(4\pi GN_5)\right.\nonumber\\
&&     - (1-\pi _X\pi_Y) 
     {\overline\Delta}_{l \alpha^{(+ 2)}}^X(k) 
     {\Delta}_{l \beta^{(+ 2)}}^Y(k) 
     k^{-1}\sin (4\pi GN_5)\cos (4\pi GN_5)\nonumber\\
 && -   (1-\pi _X\pi_Y)  
     {\overline\Delta}_{l \beta^{(+ 2)}}^X(k) 
     {\Delta}_{l \alpha^{(+ 2)}}^Y(k) 
     k^{-1}\sin (4\pi GN_5)\cos (4\pi GN_5)\nonumber\\
    &&  \left.+ (1+\pi _X\pi_Y)
     {\overline\Delta}_{l \beta^{(+ 2)}}^X(k)
     {\Delta}_{l \beta^{(+ 2)}}^Y(k) 
     \cos ^2 (4\pi GN_5)\right\}\, ,
\end{eqnarray}   
\end{widetext}
where $\pi _X$, $\pi _Y$ are the parities of the transfer functions for $X$ and $Y$.

We see from eq. (\ref{newcorr}) that the correlations are $2\pi$-periodic in 
\begin{equation}\label{thetadef}
\theta =8\pi G N_5\, ,
\end{equation} 
with purely conventional, beta-mode, contributions for $\theta$ an even multiple of $\pi$ and purely alpha-mode contributions for odd multiples.  The alpha--beta cross-contributions come from mixed parities of $X$ and $Y$, whereas the pure alpha--alpha and beta--beta contributions come from like parities.

It will be convenient to consider the cases of matched-parity and opposite-parity correlations separately.  For matched parities, we may write
\begin{equation}\label{corrmatched}
  C^{XY}_\ell = C^{XY}_{\ell\beta\beta} (1+\cos\theta )
    +C^{XY}_{\ell\alpha\alpha} (1-\cos\theta) \text{\ \ for\ \ } \pi _X=\pi _Y
\end{equation}
with
\begin{eqnarray}
C^{XY}_{\ell\alpha\alpha} &=&(2\l+1)^{-1}\int d^3\bk \,{\mathcal P}(k)  
  {\overline\Delta}_{l \alpha^{(+ 2)}}^X(k)
     {\Delta}_{l \alpha^{(+ 2)}}^Y(k) k^{-2}\qquad\\
  C^{XY}_{\ell\beta\beta} &=&(2\l+1)^{-1}\int d^3\bk \,{\mathcal P}(k)  
  {\overline\Delta}_{l \beta^{(+ 2)}}^X(k)
     {\Delta}_{l \beta^{(+ 2)}}^Y(k) \, ,
\end{eqnarray}
and, for opposite parities,
\begin{equation}\label{corropp}
  C^{XY}_\ell = -\left( C^{XY}_{\ell\alpha\beta} 
    +C^{XY}_{\ell\beta\alpha} \right)\sin\theta\text{\ \ for\ \ }\pi _X=-\pi _Y
\end{equation}  
with
\begin{eqnarray}
C^{XY}_{\ell\alpha\beta} &=&(2\l+1)^{-1}\int d^3\bk \,{\mathcal P}(k)  
  {\overline\Delta}_{l \alpha^{(+ 2)}}^X(k)
     {\Delta}_{l \beta^{(+ 2)}}^Y(k) k^{-1}\\
  C^{XY}_{\ell\beta\alpha} &=&(2\l+1)^{-1}\int d^3\bk \,{\mathcal P}(k)  
  {\overline\Delta}_{l \beta^{(+ 2)}}^X(k)
     {\Delta}_{l \alpha^{(+ 2)}}^Y(k) k^{-1}\, .\qquad
\end{eqnarray}

It should be emphasized that the factor of $k^{-1}$ accompanying the alpha-mode transfer functions does not of itself indicate any tilt of effects towards the red and away from the blue.  A moment's reflection will show that in fact
the transfer functions for the different sorts of modes have different dimensions (since the data $\alpha ^{(\pm 2)}$, $\beta ^{(\pm 2)}$ do), and cannot be directly compared.
According to eqs. (\ref{wdatone})--(\ref{wdatfour}), a unit beta-mode will correspond to $H^{(2)}\sim (\sin (k\eta))/(k\eta)$ as $\eta\downarrow 0$, whereas a unit alpha-mode will correspond to $\sim k(\cos(k\eta )/(k\eta))$; this is the reason for the factor $k^{-1}$.  
To meaningfully compare the two sorts of functions, we should consider $\Delta ^X_{l\beta ^{(\pm 2)}}$ and $k^{-1}\Delta ^X_{l\alpha ^{(\pm 2)}}$, which means
considering
the responses of the CMB to gravitational modes whose initial data are $\sim (\sin (k\eta )/(k\eta ))$ and $\sim (\cos (k\eta ))/(k\eta )$.  

We can see that, for any fixed $k$, at sufficiently small $\eta$, the alpha-type modes will dominate.  On the other hand, the transfer functions are computed by integrating the derivative of the gravitational wave field against a spherical Bessel function but weighted by factors involving the transmittance.  The transmittance vanishes very swiftly as $\eta$ decreases.  Thus for the alpha-modes to contribute distinctive effects, we must have significantly larger gravitational waves in the epoch after the Universe is opaque.  So, roughly speaking, we may expect the alpha-mode effects to be enhanced for wave-numbers below about the inverse conformal time of recombination, corresponding to multipole indices $\lesssim 100$.  However, for larger $k$ (and $l$) values, we would expect the alpha- and beta-mode effects to be roughly comparable.  In short, we expect the alpha-mode effects are enhanced over the beta-modes ones in the red, but roughly comparable in the blue, the turn-around being of order $l\sim 100$.

We may more specifically ask what happens to the transfer functions as $k$ decreases.  For very low values of $k$ (comparable to the inverse of the conformal age of the Universe), the derivative of the gravitational wave field goes like $1/(k\eta ^2)$ for beta-modes, as opposed to $k^2\eta$ for alpha-modes.  Other factors in the integrand give a power of $k$ depending on the particular type of transfer function.  For temperature, it is $k^{l-2}$ with $l$ the multipole index, so we expect a behavior $\sim k^{l-3}$ for the alpha-mode transfer functions.  In particular, the quadrupole term will diverge as $k^{-1}$ for low $k$, whereas the others will be finite.

These expectations are confirmed by numerical computations of the transfer functions and of their correlations.

\subsection{Numerical results}

Numerical computations of the transfer functions and their correlations were made with a slightly modified CLASS program, version 2.4.3~\cite{CLASSII,LegourgesTram2013}.  The cosmological data were taken from the version supplied as default, and are compatible with the WMAP/Planck 2013 results; the tensor-to-scalar ratio was $r=.35$ (with pivot scale $.05\, {\rm Mpc}^{-1}$), with primordial spectra compatible with slow-roll inflation.  Because of the slow-roll condition, the shape of the primordial spectrum depends somewhat on $r$.  
The results for the computations here are, however, close to linear in $r$, except occasionally for the $l=2$ values. 
Since the $l=2$ correlations are most susceptible to cosmic variance, a detailed catalog of their variation with $r$ is not given.

The next subsection outlines some technical details of the computation, and can be skipped.  One interesting physical point is uncovered, there, however:  the alpha-modes probe the tight-coupling regime significantly better than do the beta-modes, so an observation of alpha-mode effects would give us information about an earlier period.

\begin{figure}
\includegraphics[width=.85\linewidth]{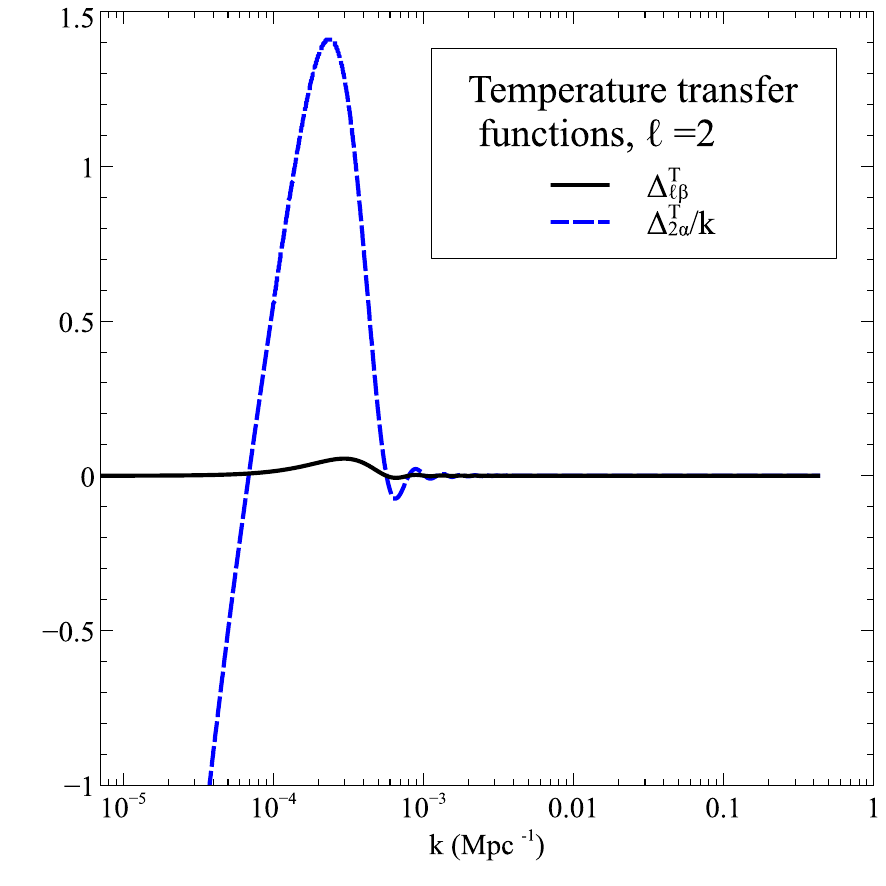}
\includegraphics[width=.85\linewidth]{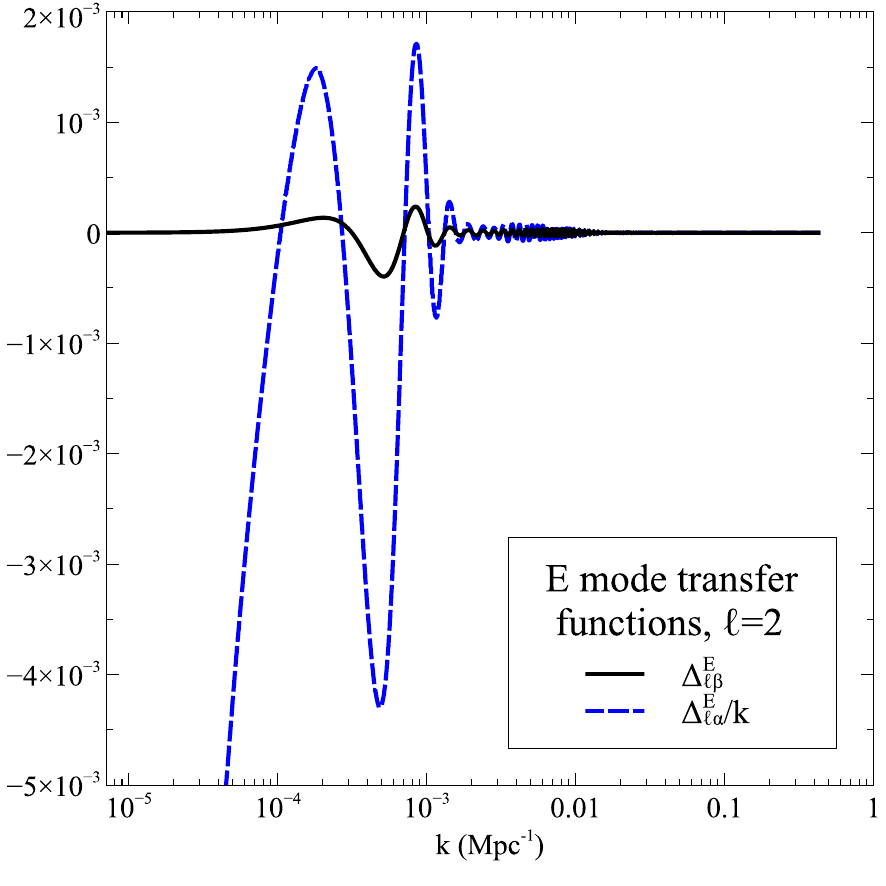}
\includegraphics[width=.85\linewidth]{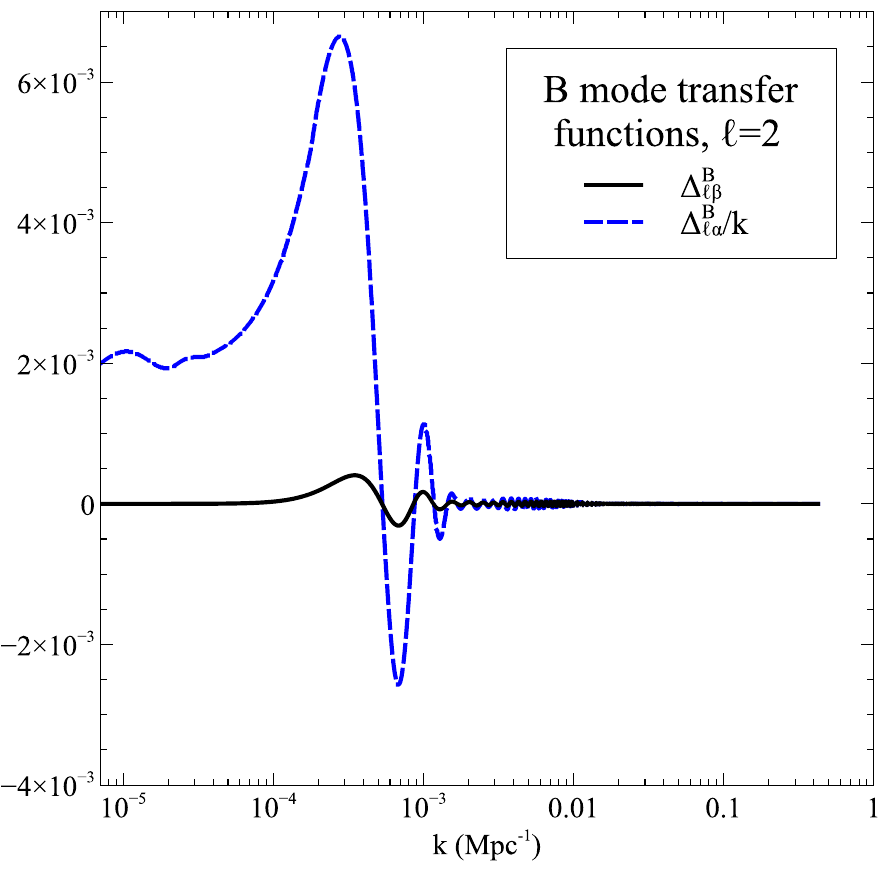}
\caption{Alpha and beta mode gravitational-wave transfer functions, computed with a modified version of CLASS, using cosmological data are compatible with Planck/WMAP 2013 results.
(Top:  temperature transfer functions.  Middle:  E mode transfer functions.  Bottom:  B mode transfer functions.  (All are for $\ell =2$.)
\label{Ttransl2fig}}
\end{figure}

\subsubsection{Computational details}

CLASS is designed so that alpha-mode waves could in principle be incorporated along with beta-modes.  However, rather than systematically rewrite the program, the initial conditions for the waves were changed when alpha-modes were required, and in each case the transfer functions written out to a file.  

Two other points are worth mentioning:  

\begin{enumerate}
\item
The code does not actually evolve the perturbations from conformal time zero, but from a suitably small initial time $\eta _{\rm init}$.  However, the code as written assumes that the data (for the beta-modes) are given in terms of the limiting value for $H^{(+2)}$ as $\eta\downarrow 0$.  This potentially leads to an error of order $(k\eta _{\rm init})^2$ for small $k\eta _{\rm init}$; the code was modified to take into account the data (for alpha- or beta-modes) being given at finite $\eta _{\rm init}$.

\item
At very early times, one is in the tight-coupling regime.  
The perturbation equations are then stiff.  The code can integrate these, but does use a tight-coupling approximation.  However, while for scalar modes a sophisticated higher-order approximation is used, for tensor modes it is 
a first-order approximation.  
For beta modes, with default precision parameters, this is adequate, but for the alpha modes it was found not to be so.  This is not really surprising, since the CMB responds to the time-derivative of the metric perturbation, and this is large for the alpha-modes at early times, whereas the beta-mode contributions are small then.  To deal with this,
in lieu of writing a higher-order tight-coupling treatment, the trigger for the start of the tight-coupling approximation was pushed downwards until stable numerical results were achieved. 
The precision parameters used were:
$\mathtt{q\_logstep\_spline=5}$,
$\mathtt{l\_linstep=20}$,
$\mathtt{start\_sources\_at\_tau\_c\_over\_tau\_h}$\allowbreak${\mathtt =}${}\allowbreak$\mathtt{1.e-3}$,
$\mathtt{tight\_coupling\_trigger\_tau\_c\_}$\allowbreak$\mathtt{over\_tau\_k=5.e-4}$.

\end{enumerate}

Physically, this means that the alpha-mode effects probe the tight-coupling regime better than do the beta-mode ones, and so a detection of alpha-mode effects would potentially give us new information.

\subsubsection{The transfer functions}

Figure \ref{Ttransl2fig} shows the alpha- and beta-mode $l=2$ transfer functions $\Delta ^X_{l\alpha} /k$ and $\Delta ^X_{l\beta}$.  One can see in these the enhanced contribution of the alpha-modes at low values of $k$ (but comparable behavior at larger $k$), and indeed the tail effects as $k\downarrow 0$, as discussed earlier.  For higher multipoles, one does not have tails, but still enhancement of the alpha-modes at low $k$.

\subsubsection{Correlations}

\begin{figure*}
\begin{minipage}[t]{.48\linewidth}
\includegraphics[width=.9\linewidth]{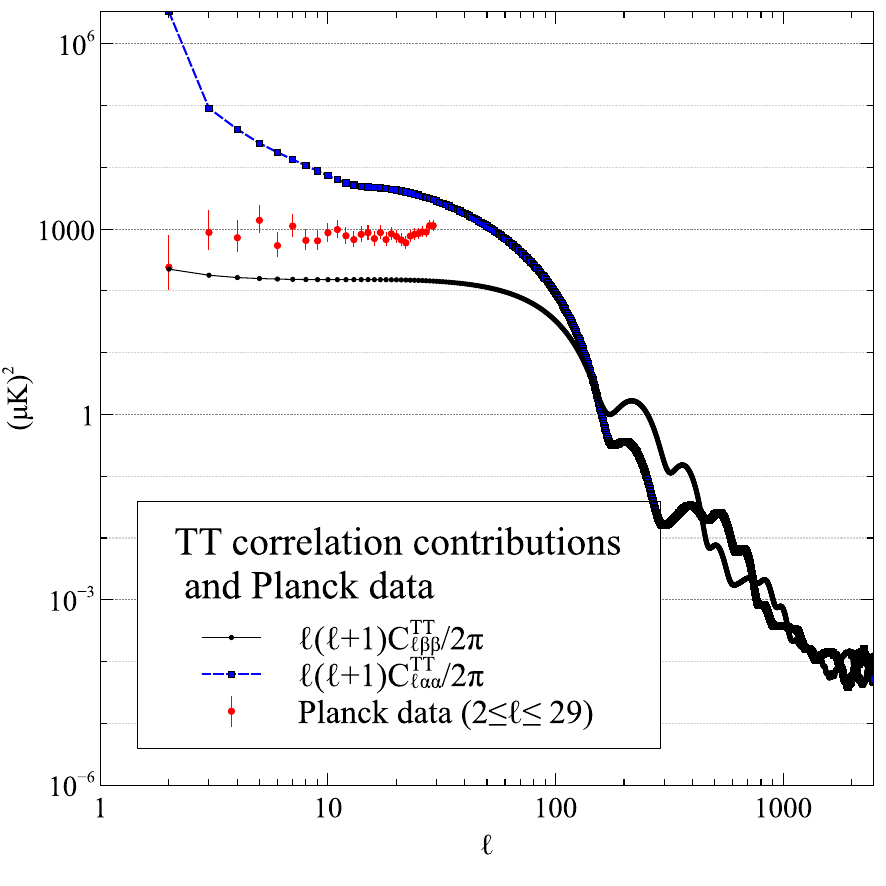}
\end{minipage}\hfill
\begin{minipage}[t]{.48\linewidth}
\includegraphics[width=.9\linewidth]{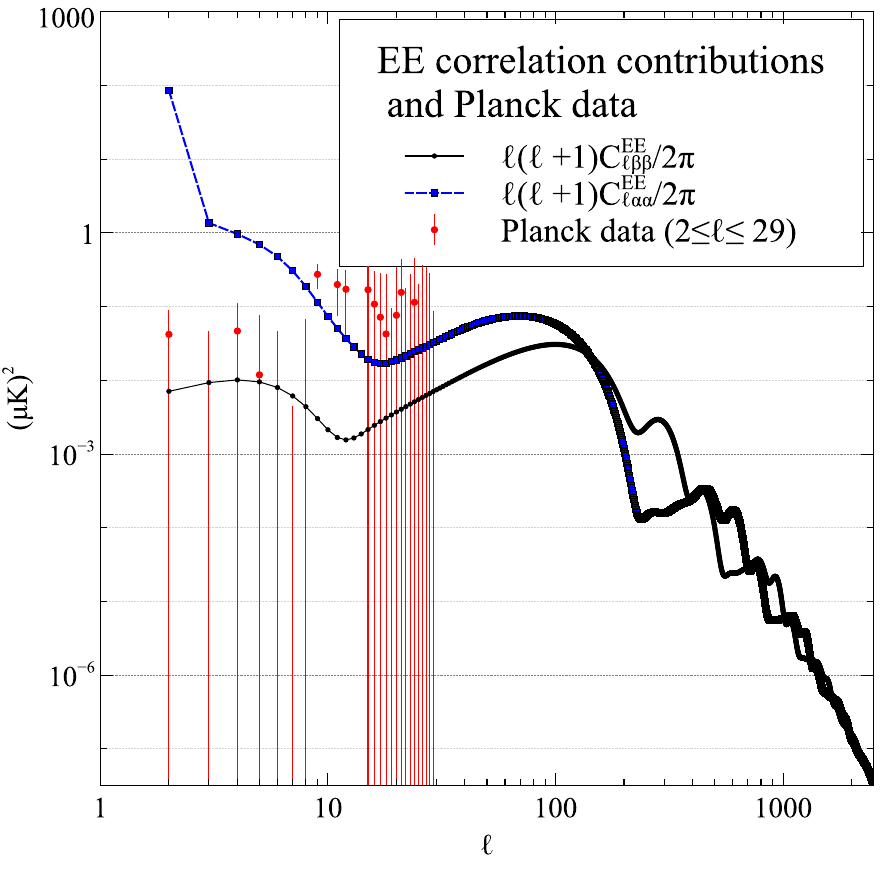}
\end{minipage}
\begin{minipage}{.48\linewidth}
\includegraphics[width=.9\linewidth]{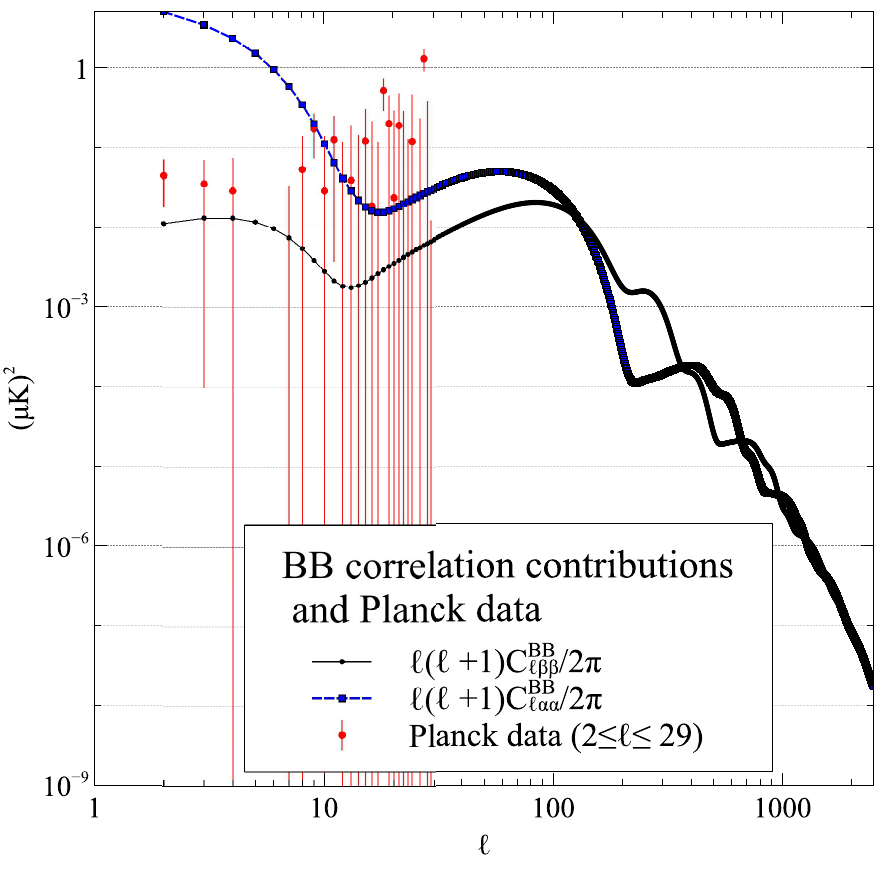}
\end{minipage}
\begin{minipage}{.48\linewidth}
\includegraphics[width=.9\linewidth]{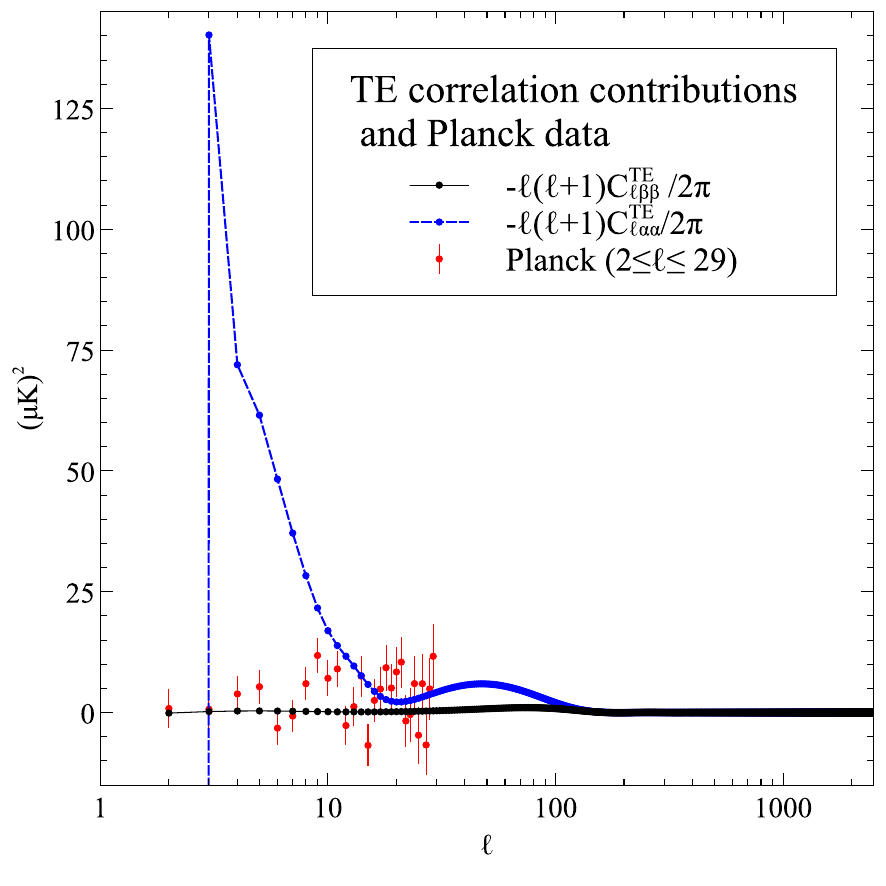}
\end{minipage}
\caption{Alpha and beta mode contributions to the TT, EE, TE and BB correlations (clockwise from top left), for a tensor-to-scalar ratio $r=.35$ and
cosmological data compatible with Planck/WMAP 2013 results, along with low-multipole data from {\em Planck}.}
\label{TTgraphfig}
\end{figure*}

\begin{figure}
\includegraphics[width=.9\linewidth]{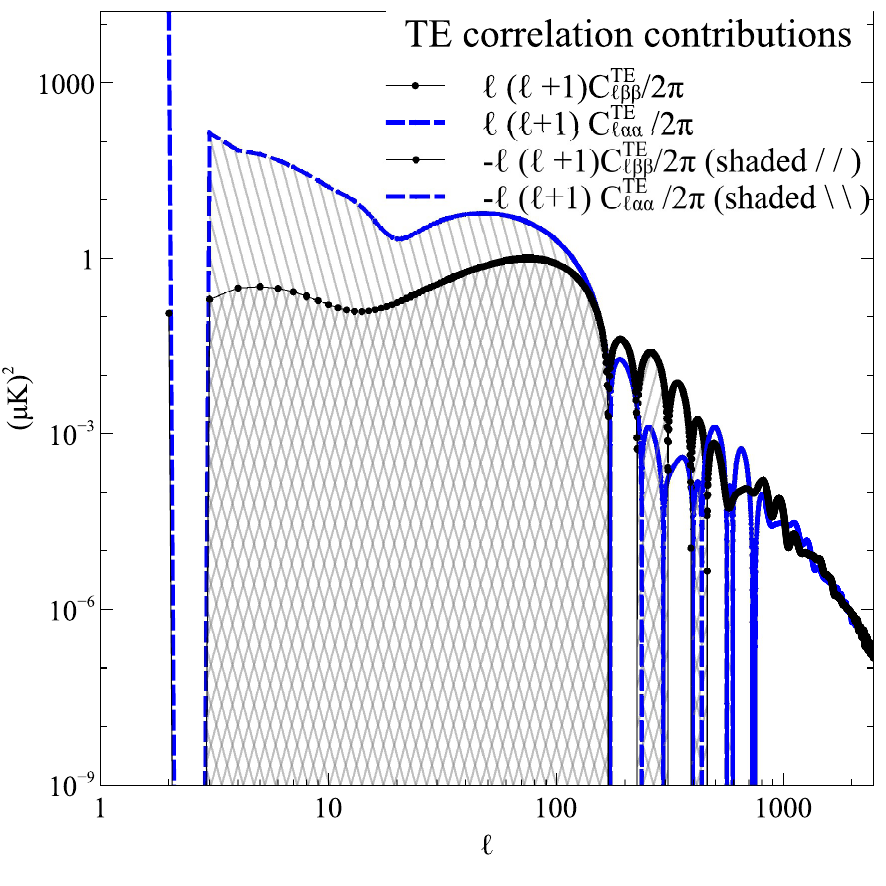}
\caption{Alpha and beta mode contributions to the TE correlations, for a tensor-to-scalar ratio $r=.35$ and
cosmological data compatible with Planck/WMAP 2013 results.
For negative correlations, the absolute value is shown but the graph is shaded below.  The signs of the alpha- and beta-mode contributions correlate closely.}
\label{TEgraphfig}
\end{figure}

Figures \ref{TTgraphfig}, \ref{TEgraphfig} show the contributions to the same-parity correlation functions from the alpha and beta modes, along with 2015 data from {\em Planck}.\footnote{The CLASS convention for the sign of $E$ is used for the quantities computed in this paper.  Since the convention for the {\em Planck} data is opposite, a minus sign appears when correlations involving one E mode are compared.}  
Recall here that the beta- and alpha-mode contributions, shown, must be combined with factors $1\pm \cos\theta$ to get the net effect (eq. (\ref{corrmatched})).
Since the correlations are compatible with the data for $\theta =0$ (modulo $2\pi$) but no firm evidence exists at present for tensor modes, one can use these results with observations to constrain $r$ and $\theta$ (modulo $2\pi$) to be close to zero.  A full investigation of this would be a large project, and will not be undertaken here.  However, some rough estimates of upper bounds of the magnitudes of the effects will be made.  

There are two issues to bear in mind in forming these estimates.  The first is cosmic variance, because the potential discrepancies between the data and the effects posited here are greatest at low multipoles.  (In particular, fluctuations due to scalars and tensors could combine with like or opposite signs.)  The second issue is that the observations
are highly correlated and the individual reported error bars, based on Fisher matrices, can only be taken as a general guide to the size of the uncertainties.

For the TT correlations, the alpha-mode factors $C^{\rm{TT}}_{l\alpha\alpha}$ are significantly higher than observations for low multipoles.  If we take the most extreme case, the quadrupole, the observed value is somewhat low compared to what is predicted on the basis of scalar perturbations; this could be due to a scalar fluctuation, a flaw in the theory, or a cancellation of a scalar and a tensor fluctuation.  In order to get an estimate, we must suppose the theory is correct.  To estimate an upper bound, let us ignore the possibility of scalar contributions and use the upper range of the reported error bar, about $800\, (\mu {\rm K})^2$.  Then we get $(r/.35)\sin ^2\theta /2 \lesssim 3\times 10^{-4}$.
The corresponding estimate from the octopole would be $(r/.35)\sin ^2\theta /2\lesssim 4\times 10^{-2}$.
Similar (or weaker) constraints come from the EE, BB and TE correlations.  

I should emphasize that these computations represent only very rough estimates for upper bounds on the magnitudes of the alpha-mode effects.  Still, it is notable how weak those estimates are.  Taking $r=.1$, the quadrupole estimate gives us $|\theta |\lesssim .05$ (modulo $2\pi$).  The octopole estimate, with $r=.1$, would give $|\theta |\lesssim .8$ (modulo $2\pi$).   If $r=.01$, the quadrupole gives us $|\theta |\lesssim .2$ (modulo $2\pi$), and the octopole gives no restriction at all.


\begin{figure*}
\begin{minipage}[t]{.48\linewidth}
\includegraphics[width=.9\linewidth]{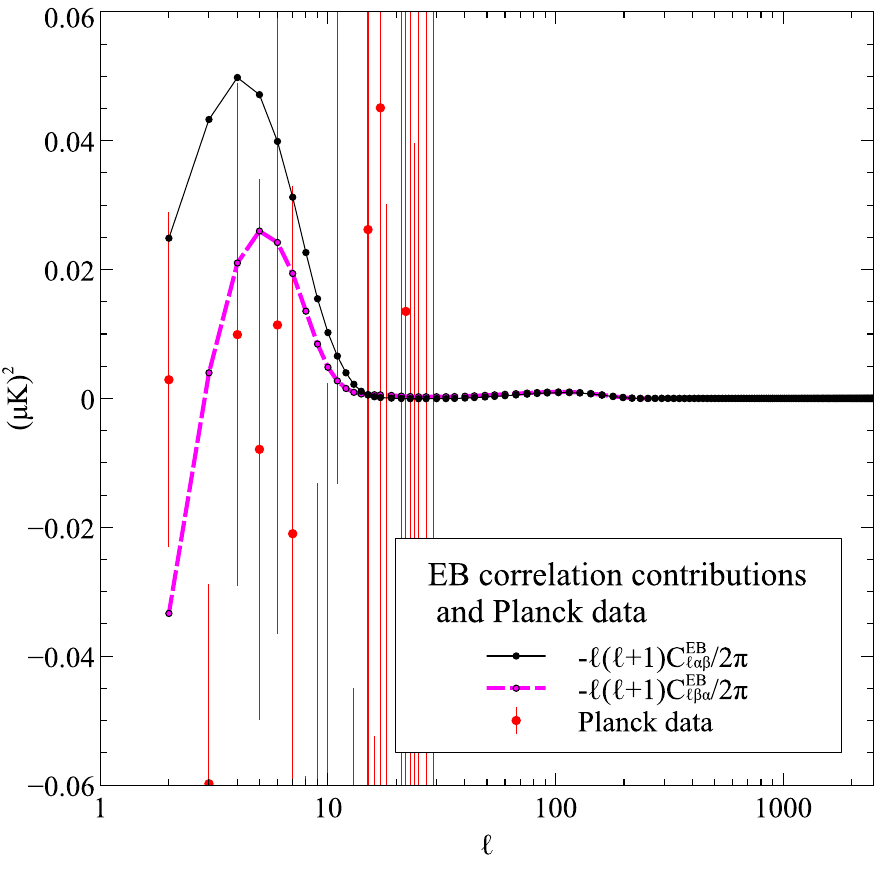}
\end{minipage}
\begin{minipage}[t]{.48\linewidth}
\includegraphics[width=.9\linewidth]{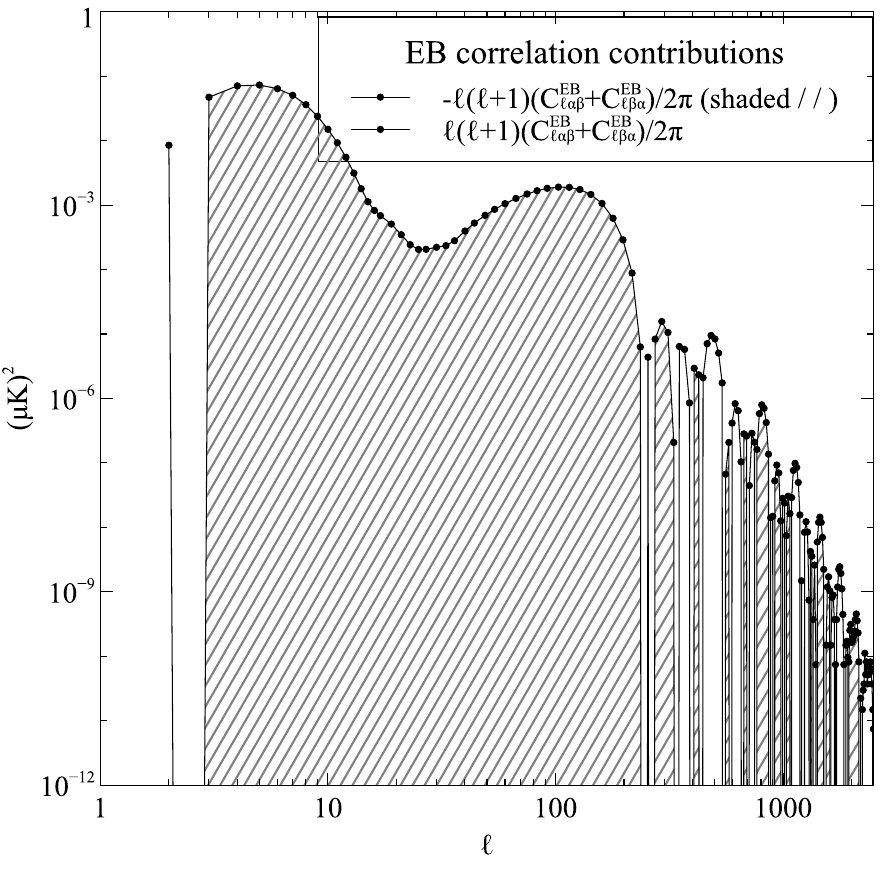}
\end{minipage}
\begin{minipage}[t]{.48\linewidth}
\includegraphics[width=.9\linewidth]{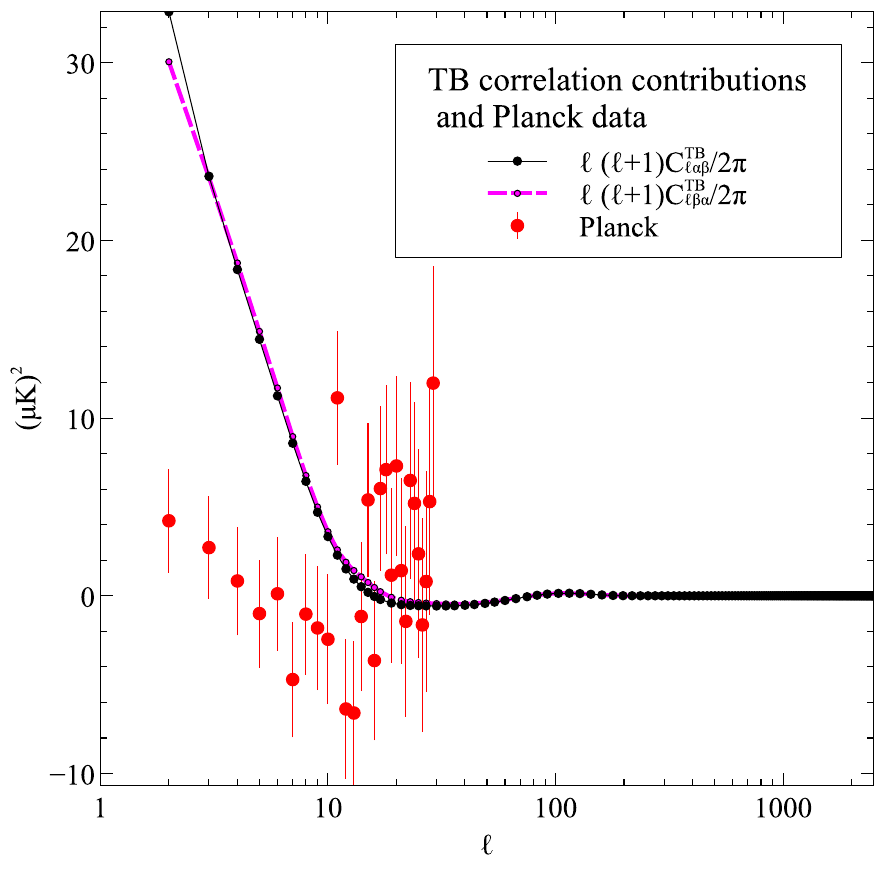}
\end{minipage}
\begin{minipage}[t]{.48\linewidth}
\includegraphics[width=.9\linewidth]{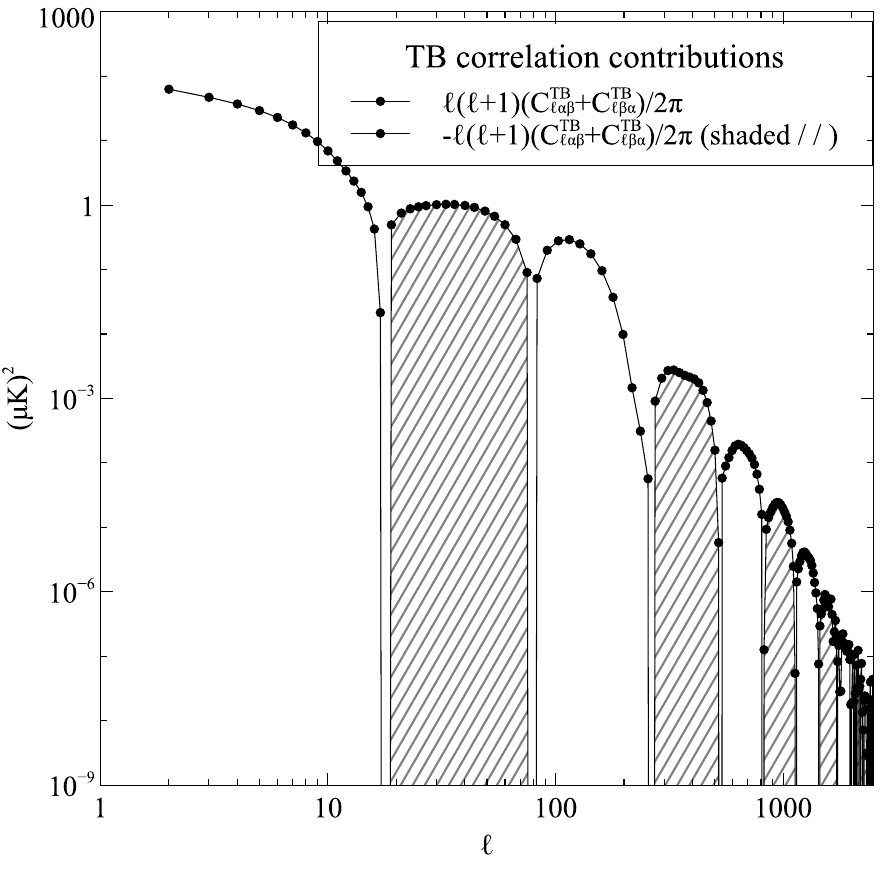}
\end{minipage}
\caption{Alpha and beta mode contributions to the EB and TB correlations, for a tensor-to-scalar ratio $r=.35$ and
cosmological data compatible with Planck/WMAP 2013 results.
Top left:  individual alpha-beta and beta-alpha cross-term contributions to the EB correlations.  Top right:  total cross-term EB contributions.  Bottom left:  individual alpha-beta and beta-alpha cross-term contributions to the TB correlations.  Bottom
right:  total cross-term TB contributions.}
\label{EBTBgraphfig}
\end{figure*}

The cross-correlations (TE, mentioned already, and EB, TB) have some interesting structure with respect to the alpha and beta modes:  the signs of the alpha-alpha and beta-beta contributions to TE are closely correlated (Figure~\ref{TEgraphfig}).
For the opposite-parity correlations EB and TB, only the sums $C^{XY}_{\ell\alpha\beta}+C^{XY}_{\ell\beta\alpha}$ would be observable.  However, in 
Figure~\ref{EBTBgraphfig}, 
I have shown the summands because of their curious behavior:  the EB graphs quite similar, and the TB ones nearly identical.  It would be worthwhile developing a detailed physical understanding of this, perhaps along the semiquantitative lines of Baskaran et al \cite{BGP}.

No reliable observational detection of EB or TB correlations has presently been considered to have been made.  In the case of EB modes, the effects described here are within the accepted observational bounds.  
On the other hand, the TB predictions are affected.  We get a bound $(r/.35)|\sin\theta|\lesssim 1/10$, corresponding to $|\theta|\lesssim .4$ for $r=.1$.  For $r=.01$, there is no restriction on $\theta$.

\section{Inhomogeneous chiral densities}

I now  consider the effects of inhomogeneous chiral densities.  These are technically more difficult to compute, and I will only consider effects of first order in the chiral density.  The main effect is to smear out the effects of the gravitational waves in wave-vector space, so that correlations are due to pairs of waves with slightly different wave-vectors.  (This is not surprising, as the waves are scattering off the chiral density.)  One new feature that occurs in this case is an expectation of circularly polarized gravitational waves.  (Circular polarization is not induced in the CMB, because the only mechanism in the model for inducing CMB polarization is Thomson scattering, which does not give circular effects.)

For simplicity, I shall take the chiral current to be localized at a conformal time $\eta _5$ corresponding to a physical time $t_5$, and write
\begin{equation}
  j_5^a =N_5\delta (t-t_5) =N_5a^{-1}\delta (\eta -\eta _5)\, .
\end{equation}
Since the treatment is linear one can simply superpose such results to get the effect more generally; however, the results as given are of some interest, as in some early-Universe models one would have as a first approximation an impulsive creation of chiral density, cut off either by local physics or expansion.

After the waves have passed through the chiral density, the functions $\lambda$ and $\rho$ will pick up, according to
eqs. (\ref{tranintphi}), (\ref{tranintpsi}), first-order changes given by
\begin{eqnarray}
  \delta _5\lambda &=&-\left( 4\pi i G\int l\cdot j_5\, ds\right)\lambda\\
  \delta _5\rho    &=&+\left( 4\pi i G\int l\cdot j_5\, ds\right)\rho\, .
\end{eqnarray}  
We
may write $\int l\cdot j_5\, ds=\int j_5\cdot d\gamma$, where $\gamma$ is a null geodesic with tangent $l^a$; the second form of the integral is manifestly independent of the parameterization of the geodesic.
If $(\eta , \bx )$ are the coordinates of a point in space--time, then the null geodesic starting from that point and directed along $l^a$ will be $\gamma (u)=(\eta +u,\bx+u{\hat\bk})$ in a convenient, but not affine, parameterization.  Then 
\begin{equation}
 \int _{-\infty}^0 N_5(\bx+u{\hat\bk})
  \delta (t-t_5)\, a\, du
= N_5(\bx+(\eta _5-\eta){\hat\bk}) \, ,
\end{equation}
using $a\, du = a\, d\eta =dt$ along $\gamma$.

The form of the results at this point is quit simple, but this is in part because they are in a mixed direct space--Fourier space representation, the argument of $N_5$ involving both the comoving coordinate $\bx$ and the Fourier direction $\hat\bk$.  To connect with the usual formalism, we must complete the passage to Fourier space.

\subsection{Fourier space}

The leading,
order-$(-i\omega )^0$, term in the metric perturbation is
\begin{eqnarray}\label{pertform}
\delta g_{ab}&\sim& \int d^3\bk\, e^{-i(k\eta -\bk\cdot\bx)}
\left( \phi (\bk ,\eta )m_am_b+\psi (\bk ,\eta ){\overline m}_a{\overline m}_b
\right)\nonumber\\
  &&+\,\text{conjugate}\nonumber\\
&=&\int d^3\bk\, e^{i\bk\cdot\bx}\times\nonumber\\
&&\left[ 
\left( e^{-ik\eta}\phi (\bk ,\eta )+e^{ik\eta}\overline\phi (-\bk ,\eta)\right) m_am_b(\bk )\right.\nonumber\\
&&\left.+\left(e^{-ik\eta}\psi (\bk ,\eta )+e^{ik\eta}\overline\psi (-\bk , \eta)\right){\overline m}_a{\overline m}_b(\bk )
\right]
\, ,
\end{eqnarray}
where the terms in square brackets give the spatial Fourier transform, as usual.  In the era before the current, this form applies with $\phi (\bk ,\eta )$, $\psi (\bk ,\eta )$ given by eq. (\ref{phipsiinit}).  After that, however, there will be an additional contribution to eq. (\ref{pertform}) given by
\begin{eqnarray}
\delta_5 g_{ab}&\sim& 4\pi iG\int d^3\bk\, e^{-i(k\eta -\bk\cdot\bx)}a^{-1}\times\\
&&
\left( -\lambda (\bk) N_5(\bx+(\eta _5-\eta)\hat\bk)m_am_b
\right.\nonumber\\
&&\left.+\rho (\bk )N_5(\bx+ (\eta _5-\eta)\hat\bk){\overline m}_a{\overline m}_b
\right)+\,\text{conjugate}\, .\nonumber
\end{eqnarray} 
The idea now will be to Fourier transform this with respect to the spatial variables, in order to derive a scattering formula for the gravitational waves in wave-vector space.  In principle, there are various technical complications which can arise in such a calculation (related to gauge freedom), but we shall see that in the high-frequency limit there is a simple result.

The spatial Fourier transform is
\begin{widetext}
\begin{eqnarray}\label{metfive}
(2\pi )^{-3}\int \delta _5 g_{ab}e^{-i{\acute\bk}\cdot\bx}d^3\bx
&=&4\pi iGa^{-1}\int d^3\bk\,
  \left[ e^{-ik\eta}
{\widehat N_5}({\acute\bk}-\bk) 
  e^{i{\hat\bk}\cdot({\acute\bk}-\bk)(\eta _5-\eta)}\left( -\lambda (\bk)
m_am_b +\rho (\bk){\overline m}_a{\overline m}_b\right)   \right.\nonumber\\
  &&\left. 
  +e^{ik\eta}\overline{\widehat N_5}({\acute\bk}+\bk) 
e^{-i{\hat\bk}\cdot({\acute\bk}+\bk)(\eta _5-\eta)}\left( -{\overline\lambda} (\bk)
{\overline m}_a{\overline m}_b +{\overline\rho} (\bk){m}_a{m}_b\right) \right] \nonumber\\
&=&4\pi iGa^{-1}\int d^3\bk\,    
  \left[ e^{-i{\acute k}\eta}
{\widehat N_5}({\acute\bk}-\bk) 
  e^{i{\hat\bk}\cdot{\acute\bk}(\eta _5-\eta)+i{\acute k}\eta -ik\eta _5)}\left( -\lambda (\bk)
m_am_b +\rho (\bk){\overline m}_a{\overline m}_b\right)  \right.\nonumber\\ 
  &&\left. 
+e^{i{\acute k}\eta}\overline{\widehat N_5}({\acute\bk}+\bk) 
e^{-i{\hat\bk}\cdot{\acute\bk}(\eta _5-\eta)-i{\acute k}\eta +ik\eta _5)}\left( -{\overline\lambda} (\bk)
{\overline m}_a{\overline m}_b +{\overline\rho} (\bk){m}_a{m}_b\right) \right] \nonumber\\
&=&4\pi iGa^{-1}\int d^3\bk\,  
  \left[ e^{-i{\acute k}\eta}
{\widehat N_5}({\acute\bk}-\bk) 
 e^{i({\acute\bk}-{\hat\bk}\cdot{\acute\bk})\eta +i({\hat\bk}\cdot{\acute\bk}-k)\eta _5}
  \left( -\lambda (\bk)
m_am_b +\rho (\bk){\overline m}_a{\overline m}_b\right)\right.\nonumber\\ 
  && \left. 
+e^{i{\acute k}\eta}\overline{\widehat N_5}({\acute\bk}+\bk) 
e^{i({\hat\bk}\cdot{\acute\bk}-{\acute k})\eta +i(k-{\hat\bk}\cdot{\acute\bk})\eta _5}
\left( -{\overline\lambda} (\bk)
{\overline m}_a{\overline m}_b +{\overline\rho} (\bk){m}_a{m}_b\right) \right] \, ,
\end{eqnarray}
\end{widetext} 
where $\widehat N_5$ is the Fourier transform of $N_5$.  In each case, here, the polarization factors $m_am_b$, ${\overline m}_a{\overline m}_b$ are taken at $\bk$.

Eq. (\ref{metfive}) represents a single Fourier mode of the perturbation, and should therefore have the same form as the 
factor in square brackets in eq. (\ref{pertform}), for some different $\phi$ and $\psi$ and with the variable $\bak$ in eq. (\ref{metfive}) corresponding to $\bk$ in eq. (\ref{pertform}).
This does not appear to be the case, for two reasons.  First, since in this regime there is no chiral current, one would think that the $\eta$-dependence within the square brackets would simply be through factors $e^{\pm i{\acute k}\eta}$ (modulo gauge issues), but other $\eta$-dependent factors appear as well.  Second,
the polarizations $m_am_b(\bk )$, ${\overline m}_a{\overline m}_b(\bk )$ in eq. (\ref{metfive}) ought to be recast in terms of $m_am_b(\bak )$, ${\overline m}_a{\overline m}_b(\bak )$ (also modulo gauge).  These two concerns do not cancel each other out.  The resolution is different.

The issue is that we are applying the high-frequency limit not to a single wave but to a distribution of them weighted by $\lambda (\bk )$ and $\rho (\bk )$, and some sort of uniformity in the expansion must be assumed.  
This enters in the expressions above in the factors ${\widehat N_5}(\bak \mp \bk )$.  Since the basic assumption of the high-frequency limit is that the gravitational waves vary more rapidly than do quantities like the chiral density, the Fourier transform $\widehat N_5$ cannot be supported significantly for very large wave-vectors; it must fall off.  But since $\bk$ is very large in the high-frequency expansion, we must have $\bak$ relatively close to $\pm \bk$.  

To proceed quantitatively, we will assume $\|\bak\mp\bk\|$ is $O(1)$ in $\|\bk\|$ where ${\widehat N_5}(\bak\mp\bk )$ has significant support.  Then (taking the upper sign, for simplicity)
we have
\begin{eqnarray}
 \|\bak \| &=& \| \bk +(\bak - \bk) \| \nonumber\\
   &=&\|\bk\| +{\hat\bk}\cdot (\bak -\bk ) +O(\|\bk\|^{-1})\nonumber\\
  &=& {\hat\bk}\cdot\bak+O(\|\bk\|^{-1})\label{firstest}\\
\|\bak\| -{\hat\bk}\cdot\bak &=&O(\|\bk\|^{-1})\label{secondest}\\
\| \hat{\bak} -\hat\bk\|  &=&\sqrt{2- 2{\hat\bak}\cdot\hat\bk }
=O(\|\bk\| ^{-1})\, ,\label{thirdest}
\end{eqnarray}
where the last line follows from Eq. (\ref{firstest}).  Equations (\ref{firstest}, \ref{secondest}) allow us to simplify the exponents in eq. (\ref{metfive}); Equation (\ref{thirdest}) justifies neglecting, to leading order, discrepancies in polarizations relative to $\bk$ versus $\bak$.
With these simplifications, the formula (\ref{metfive}) becomes
\begin{eqnarray}
&&(2\pi )^{-3}\int \delta _5 g_{ab}e^{-i{\bak}\cdot\bx}d^3\bx
=
4\pi iGa^{-1}\int d^3\bk\,\times\nonumber\\
&&\quad\left[ e^{-i{\acute k}\eta}
{\widehat N_5}(\bak -\bk) 
 e^{i(\acute k -k)\eta _5}
  \left( -\lambda (\bk)
m_am_b +\rho (\bk){\overline m}_a{\overline m}_b\right) \right.
  \nonumber\\
  &&\left. 
  +e^{i{\acute k}\eta}\overline{\widehat N_5}({\bak}+\bk ) 
e^{i(k-\acute k )\eta _5}
\left( -{\overline\lambda} (\bk)
{\overline m}_a{\overline m}_b +{\overline\rho} (\bk){m}_a{m}_b\right) \right] \, .\qquad
\end{eqnarray} 
Thus the effect is (appropriately relabeling)
\begin{eqnarray}
\delta _5\lambda (\bk )&=&-
4\pi iG\int d^3\bak \, {\widehat N}_5(\bk-\bak ) e^{i(k-\acute k )\eta _5}\lambda (\bak )\label{dlam}\\
\delta _5\rho (\bk)&=&+
4\pi iG\int d^3\bak\,  {\widehat N}_5(\bk-\bak ) e^{i(k-\acute k )\eta _5}\rho (\bak )\, .\label{drho}
\end{eqnarray} 
Evidently the effect of the chiral density is to smear out the wave-packets defined by the distributions $\lambda (\bk )$, $\rho (\bk )$  by an amount $\sim N_5(\bx -(\eta -\eta _5)\hat\bk )$.

In turn, this effect is the same as would be achieved by a perturbation of the initial data in the amount
\begin{eqnarray}
\delta _5\alpha ^{(2)} &=&\sqrt{2/3} (\delta _5\rho (\bk ) +\delta _5\overline\rho (-\bk ))\nonumber\\
&=&\sqrt{2/3}(4\pi iG)\left( \int d^3\bak \,{\widehat N}_5(\bk -\bak ) e^{i(k-\acute k )\eta _5} \rho (\bak ) \right.\nonumber\\
  &&\left.
  -\int d^3\bak\, \overline{\widehat N}_5 (-\bk -\bak ) e^{-i(k-\acute k )\eta _5} \overline\rho (\bak ) \right) \nonumber\\
  &=&(2\pi iG)\left( \int d^3\bak \,{\widehat N}_5(\bk -\bak ) e^{i(k-\acute k )\eta _5} (\alpha (\bak )-\beta (\bak )/i\acute k) \right.\nonumber\\
  &&\left.
  -\int d^3\bak\, \overline{\widehat N}_5 (-\bk -\bak ) e^{-i(k-\acute k )\eta _5} (\overline\alpha (\bak )+\overline\beta (\bak )/i\acute k) \right) \nonumber\\
&=&-4\pi  G\int \frac{d^3\bak}{\acute k}\, {\widehat N}_5(\bk -\bak ) \cos ((k-\acute k )\eta _5) \beta ^{(2)}(\bak ) \label{changeone}
\end{eqnarray}  
\begin{eqnarray}
\delta _5\alpha ^{(-2)} &=&4\pi G\times\\
&&  \int \frac{d^3\bak}{\acute k} \,{\widehat N}_5(\bk -\bak ) \cos ((k-\acute k )\eta _5) \beta ^{(-2)}(\bak )\nonumber \\
\delta _5\beta ^{(\pm 2)}&=&\mp 4\pi G\times  \label{changefour}\\  
&& k\int \frac{d^3\bak}{\acute k}\, {\widehat N}_5(\bk -\bak )\sin ((k-\acute k) \eta _5) \beta ^{(\pm 2)}(\bak )\, .\nonumber
\end{eqnarray}
From these and eqs. (\ref{coreone}), (\ref{coretwo}) we have
the expectations
\begin{widetext}
\begin{eqnarray}
\langle\delta _5\overline\alpha ^{(\pm 2)}(\bk ) \beta ^{(\pm 2)}(\bak )\rangle
  &=& \mp 4\pi G{\widehat N}_5(-\bak -\bk )\cos((k-\acute k)\eta _5){\acute k}^{-1}{\cal P}(\acute k )
   \label{grcorone}\\
\langle\overline\beta ^{(\pm 2)}(\bk )\delta _5\alpha ^{(\pm 2)}(\bak ) \rangle
&=&\mp 4\pi G{\widehat N}_5(\bak +\bk ) \cos ((k-\acute k )\eta _5) k^{-1}{\cal P}(k)\\ 
\langle 
    \delta _5\overline\beta ^{(\pm 2)}(\bk ) \beta ^{(\pm 2)}(\bak )\rangle
    &=&\mp 4\pi G{\widehat N}_5(-\bak -\bk )\sin((k-\acute k )\eta _5)) {\acute k}^{-1}{\cal P}(\acute k)\\
\langle  \overline\beta ^{(\pm 2)}(\bk ) \delta _5\beta ^{(\pm 2)}(\bak )\rangle
  &=&\mp 4\pi G\acute k {\widehat N}_5(\bak +\bk )\sin ((\acute k - k)\eta _5) k^{-1}{\cal P}(k)\, .
  \label{grcorfour}
\end{eqnarray} 

\subsection{Effect on CMB correlations}

With the previous subsection's results, we may immediately compute the change
in the CMB correlations:
\begin{eqnarray}
\delta _5C^{XY}_l&=&(2l+1)^{-1} \sum_{\pm}
  \int d^3\bk d^3\bak
  \left( {\overline\Delta}^X_{l\alpha ^{(\pm 2)}}(k)
          \Delta ^Y_{l\beta^{(\pm2)}}(\acute k )
          \langle \delta _5{\overline\alpha}^{(\pm 2)}(\bk )
              \beta^{(\pm 2)}(\bak )\rangle
              +{\overline\Delta}^X_{l\beta ^{(\pm 2)}}(k)
          \Delta ^Y_{l\alpha^{(\pm2)}}(\acute k )
          \langle {\overline\beta}^{(\pm 2)}(\bk )
              \delta _5\alpha^{(\pm 2)}(\bak )\rangle\right.\nonumber\\
              &&\left.
              +{\overline\Delta}^X_{l\beta ^{(\pm 2)}}(k)
          \Delta ^Y_{l\beta^{(\pm2)}}(\acute k )
          \langle \delta _5{\overline\beta}^{(\pm 2)}(\bk )
              \beta^{(\pm 2)}(\bak )
              +{\overline\beta}^{(\pm 2)}(\bk )
              \delta_5\beta ^{(\pm 2)}(\acute k )\rangle\right)\\
    &=&-(2l+1)^{-1}(1-\pi _X\pi _Y)4\pi G
              \int d^3\bk d^3\bak {\hat N}_5(\bak +\bk) \times\nonumber\\
&&  \left( \left({\overline\Delta}^X_{l\alpha ^{(2)}}(k)
          \Delta ^Y_{l\beta^{(2)}}(\acute k )
          {\acute k}^{-1}{\mathcal P}(\acute k)
              +{\overline\Delta}^X_{l\beta ^{( 2)}}(k)
          \Delta ^Y_{l\alpha^{(2)}}(\acute k )
          k^{-1}{\mathcal P}(k)       \right)\cos ((k-\acute k )\eta _5)
          \right.\nonumber\\
              &&\left.
              +{\overline\Delta}^X_{l\beta ^{( 2)}}(k)
          \Delta ^Y_{l\beta^{(2)}}(\acute k )
          (
          {\acute k}^{-1}{\mathcal P}(\acute k)
          - k^{-1}{\mathcal P}(k))
          \sin ((k-\acute k)\eta _5)
          \right)      \, .
\end{eqnarray}
\end{widetext}

There are four important features of this: 
(a) The only changes which are first-order in $N_5$  are for the mixed-parity correlations, as they should be (given the odd parity of $N_5$).  
(b) The correlations found here are nonlocal in Fourier space, in contrast to the standard formulas for the effects of gravitational waves.  This nonlocality is not really very surprising, however, because, roughly speaking, the waves are scattering off the chiral density, and this has the effect of smearing them.
(c) In these formulas, the conformal time $\eta _5$ at which the chiral density was encountered enters explicitly.  Therefore observation of these effects would give us information about the conformal time at which the gravitational waves passed through the chiral density.
(d) Because the transfer functions depend only on the magnitudes of the wave-vectors, they are real, and hence the correlation here responds only to the real part of ${\hat N}_5$, which is to say the symmetric (under reflection) part of $N_5(\bx )$.  This is really a consequence of parity properties.  There is an interesting contrast with gravitational waves, however, as we see next.

\subsection{Circularly polarized gravitational waves}

While the chiral current does not lead, by the effects considered here, to circular polarizations in the CMB, it does give rise to such polarizations for the gravitational waves.  More precisely, it induces non-trivial expectations of such polarizations.

A measure of the expected circular polarization of gravitational waves would be
\begin{equation}\label{graV}
 V(\bk ,\bak) =\delta _5\langle {\overline\rho}(\bk )\rho (\bak ) -
{\overline\lambda}(\bk )\lambda (\bak )
\rangle\, .
\end{equation}
The diagonal values $V(\bk ,\bk )$
may be thought of as a gravitational analog of the corresponding optical Stokes parameter.  A computation using the formulas above gives 
\begin{eqnarray}
&&3\pi i G\cos ((k-\acute k )\eta _5)\times\label{graVY}\\
 && \left( ( k)^{-2}{\widehat N}_5(\bk+\bak ){\cal P}( k )
 -(\acute k)^{-2}{\widehat N}_5(-\bk-\bak ){\cal P}(\acute k )
     \right)
    \nonumber\\
    &+&3\pi G \frac{\sin ((k-\acute k)\eta _5)}{k\acute k}
\times\nonumber\\
&&    \left( 
 (\acute k)^{-1}{\widehat N}_5(-\bk -\bak ){\cal P}(\acute k)
   -( k)^{-1}{\widehat N}_5(\bk +\bak ){\cal P}( k)\right)
     \nonumber
\end{eqnarray}
for $V(\bk ,\bak )$.  Note that this is sensitive to the odd part of $N_5$.  Also, while this two-point function is sensitive to the conformal time $\eta _5$ at which the waves pass through the conformal density, the gravitational Stokes parameter
\begin{equation}
  V(\bk ,\bk ) =   3\pi iG k^{-2}{\mathcal P}(k) 
    ({\hat N}_5(2\bk ) -{\hat N}_5(-2\bk ))
\end{equation} 
is not.

\section{Discussion}

The main result here is a remarkably clean formula for the gryotropy of gravitational waves by a chiral column density $N_5$:  a wave's polarization is rotated by $2\pi \ell _{\rm Pl}^2N_5$.  This is striking enough to suggest that there are important further connections between chiral physics and gravity, at very high energies or in the very early Universe.  Of course, such connections have been suggested for a long time; what is different here is that the primary effect does not rely on postulating any exotic physics:  it is a consequence of conventional general relativity.

This novel effect grows out of the fact that usual spin one-half stress--energy has a term depending on derivatives of the metric --- unlike the behavior of other conventional elementary fields.  This meant that the equations governing the transport of gravitational wave profiles along their geometric-optics trajectories respond to the presence of fermions. That coupling turns out to be to the chiral current $j^5_a$, but with a Planck-area prefactor.  Thus the effects involved are tiny unless very large chiral column densities $N_5$ can be achieved.

Because $j^5_a$ is parity-odd, this chiral coupling will interconvert E- and B-mode gravitational waves, and in particular a distribution of waves 
which initially had E-E or B-B correlations would develop
E-B ones after passing through a chiral density.  These would in principle be directly observable, if we were able to detect the gravitational waves.
Because of the waves' couplings to the cosmic microwave background, they would produce E-B (and T-B) correlations in the CMB as well.  

However, the passage of the gravitational waves through a chiral density will give other, arguably more primitive effects:  it will disturb the ``cosmic coherence'' which determines the subspace (what I have called the beta-modes) of initial conditions for gravitational waves within big-bang models.  After the waves pass through the chiral density, they will acquire components in a complementary subspace (the alpha-modes).
This means in particular that in order to compute the effects on the CMB we require transfer functions responding to those complementary degrees of freedom.  
These transfer functions were computed numerically with a slightly modified CLASS program,  and the resulting correlations worked out (to all orders in $N_5$ in the case where it was spatially constant; to first order in general).

In the spatially constant case, the effects were $2\pi$-periodic in $\theta =8\pi \ell _{\rm Pl}^2N_5$.  A very rough comparison with data from {\em Planck} was made, and this did not uncover any restrictions on $\theta$ one could be confident in for tensor-to-scalar ratio $r\lesssim .01$.

It is worthwhile noting that gravitational gyrotropy has implications, in principle, for gravitational memory effects.  These effects, which go back to Bondi \cite{BVM} and have become best known through the paper of Christodoulou \cite{Christodoulou1991}, concern a sort of gauge mismatch between two quiescent regimes at null infinity for an isolated system, bracketing an emission of gravitational radiation.  The mismatch is characterized by the E-mode difference in the Bondi shears of the two regimes.  However, it has long been known that in principle there may be a B-mode contribution as well.  What the work here shows is that, if the waves pass through a chiral density on their way outwards, there will be contributions to the B-mode changes in consequence of the E-mode ones.  The B-mode effects, unlike the E-mode ones, cannot be absorbed in changes of gauge; they contribute to the spin angular momentum of the gravitational radiation \cite{ADH2007}.

\acknowledgments
It is a pleasure to thank Arthur Kosowksy for his explanations of some points of the CMB computations and refer to his very clear article \cite{Kosowsky1996}.  I am also thankful for the public availability of the CLASS program, 
and to Julien Lesgourges for answering questions about it, which made clear some details of how the computations are done.  The graphs were prepared with VEUSZ.

\appendix

\section{Perturbing a Fermi field and its stress--energy}

This appendix depends on some of the formulas derived in section 2.2.

\subsection{Notation and conventions}

The conventions for space--time and two-spinors are those of Penrose and Rindler~\cite{PR1984,PR1986}. 
These books do not use four-component spinors in their analysis, and their conventions for two-component spinors and space--time are compatible with much other work.  
These books do give, in notes (see \cite{PR1984}, p. 221 and \cite{PR1986}, p. 460), formulas for converting to Dirac spinors,
but we shall {\em not} use those conventions, because they are adapted to an opposite sign for $\gamma _a\gamma_b +\gamma _b\gamma _a$ from what is generally used in the quantum-field-theoretic literature.  With our choices, our conventions for four-spinors conform to those of Schweber \cite{Schweber1961}.

The space--time metric is $g_{ab}$; it has signature $+{}-{}-{}-$.  The curvature tensors satisfy $[\nabla _a,\nabla _b]v^d =R_{abc}{}^dv^c$, $R_{ac}=R_{abc}{}^b$, $R=R_a{}^a$, $G_{ab}=R_{ab}-(1/2) Rg_{ab}$.  The volume form is $\epsilon _{abcd}$ and we have $\epsilon _{txyz}=+1$ in a right-handed future-pointing orthonormal basis.

Two-spinors are denoted by symbols like $\kappa ^A$, conjugate spinors as $\lambda ^{A'}$.  The conjugation map is denoted by an overbar: $\kappa ^A\mapsto {\overline\kappa}^{A'}$.  Spin-space is equipped with a non-degenerate skew form $\epsilon _{AB}$, whose conjugate is denoted $\epsilon _{A'B'}$.  

Spinor and vector indices are related through the Infeld--van der Waerden symbols $\sigma _a{}^{AA'}$, so that we put $v^{AA'}=v^a\sigma _a{}^{AA'}$, etc.  We have
$g_{ab}=\sigma _a{}^{AA'}\sigma _b{}^{BB'}\epsilon _{AB}\epsilon _{A'B'}$.  The spinor form of the alternating tensor $\epsilon _{abcd}$ is $i\epsilon _{AC}\epsilon _{BD}\epsilon _{A'D'}\epsilon _{B'C'}-i\epsilon _{AD}\epsilon _{BC}\epsilon _{A'C'}\epsilon _{B'D'}$.

A Dirac spinor will be represented by a pair of two-spinors:
\begin{equation}
\psi =\left[\begin{matrix} \psi ^{Q'}\\ \psi _Q\end{matrix}\right]\, .
\end{equation}
The Dirac symbols are given by
\begin{equation}
\gamma _a=\sqrt{2}\left[\begin{matrix}
  0&\sigma _a{}^{P'Q}\\ \sigma _{aPQ'}&0\end{matrix}\right]
\end{equation}
and satisfy $\gamma _a\gamma _b+\gamma _b\gamma _a=2g_{ab}$.  
As noted above, these conventions for the {\em relation} between two- and four-component spinors, and for the Dirac gammas, differ from those  of Penrose and Rindler.  With the present conventions, we have an exact correspondence between the standard basis described on pp. 120--125 of Penrose and Rindler \cite{PR1984} or pp. 6--8 of \cite{PR1986} and the Weyl basis on p. 79 of Schweber \cite{Schweber1961}.

The Dirac adjoint is
\begin{equation}
\tilde\psi =\left[\begin{matrix} {\overline\psi}_{Q'}
  &{\overline\psi}^Q\end{matrix}\right]\, ,
\end{equation}
and the Dirac current is thus
\begin{equation}
  j_{AA'}=\tilde\psi \gamma _{AA'}\psi
   =\sqrt{2}\left( {\overline\psi}_{A'}\psi _A +{\overline\psi}_A \psi _{A'}\right)\, .
\end{equation}
More correctly, we should treat $\psi$ as a Fermi field operator and use the antisymmetrized form
\begin{equation}
  (1/2)[{\tilde\psi} \gamma _{AA'},\psi ]
   =2^{-1/2}\left({\overline\psi}_{A'}\psi _A +{\overline\psi}_A \psi _{A'}-\psi _A{\overline\psi}_{A'}-\psi _{A'}{\overline\psi}_A\right)\, .
\end{equation}
This gives as usual the current of fermions minus antifermions.

We take
\begin{equation}
\gamma _5=\frac{1}{24}\epsilon _{abcd}\gamma ^a\gamma ^b\gamma ^c\gamma ^d =\left[\begin{matrix} i&\\&-i\end{matrix}\right]\, .
\end{equation}     
(Many authors use $-i$ times this.)  The projectors to the right- (respectively, left-) handed spinors are 
\begin{equation}
(1/2)\left[ 1\pm i\gamma _5\right]
=\left[\begin{matrix} 0&\\&1\end{matrix}\right]\, ,\quad
\left[\begin{matrix} 1&\\&0\end{matrix}\right]
 \, .
 \end{equation}  
We define the chiral current as
\begin{eqnarray}
  j^5_a&=&i{\tilde\psi}\gamma _a\gamma _5 \psi\nonumber\\
  &=& {\tilde\psi}\gamma _a \left[ -(1/2)(1-i\gamma _5) +(1/2)(1+i\gamma _5)\right]\psi\nonumber\\
  &=&\sqrt{2}\left( {\overline\psi}_{A'}\psi _A -{\overline\psi}_A \psi _{A'}\right)\, ,
\end{eqnarray}
which gives the right- minus left-handed fermions.
Again, strictly speaking, we should consider an antisymmetrized version of this.

Finally, we note that the charge conjugate of a spinor is
\begin{eqnarray}
  \psi _{\rm c}&=&\left[\begin{matrix}0&\epsilon ^{Q'R'}\\
    \epsilon _{QR}&0\end{matrix}\right] \left[\begin{matrix}
      {\overline\psi} ^R\\ {\overline\psi}_{R'}\end{matrix}
      \right]\nonumber\\
  &=&\left[\begin{matrix} {\overline\psi}^{Q'}\\
    {\overline\psi}_Q\end{matrix}\right]\, .
\end{eqnarray}    
(The usual basis-dependent formulas in terms of $\gamma _2$ being replaced by the matrix with epsilons.)

We turn now to the perturbations of the Fermi field and its stress--energy.  We must begin by discussing the perturbation of the spin structure; with this in hand, we can consider how the field and the stress--energy respond.

\subsection{Change in the spin structure}

When we consider a first-order perturbation $\delta g_{ab}=h_{ab}$ of the metric, we must fix a convention for how to perturb the spin-structure.  The natural choice is to take
\begin{equation}
\delta \sigma _a{}^{AA'} =\frac{1}{2}h_a{}^{AA'}\, .
\end{equation}
(Here we follow the usual practice of raising and lowering indices with respect to the background quantities --- although, since $h_{ab}$ is first-order, the ambiguity in the raising of its spinor indices is irrelevant.)
With this choice, the skew spinor $\epsilon _{AB}$ is preserved. 

The covariant derivative operator must change as well, so that the conditions 
\begin{equation}\label{concond}
\nabla _a\sigma _b{}^{BB'}=0\, ,\ \nabla _a\epsilon _{BC}=0
\end{equation} preserved (to first order).
After some algebra, one finds that the perturbation is
\begin{equation}
  (\delta\nabla _a)\kappa ^Q=\gamma _{aP}{}^Q\kappa ^P\, ,
\end{equation}
where
\begin{equation}
  \gamma _{aBC}=\frac{1}{2}\nabla _{(B|Q'|}h_{C)}{}^{Q'}{}_a\, ,
\end{equation}
as one can verify by checking eq. (\ref{concond}).
(Note that this gamma is not a Dirac symbol and in fact has a different index structure.)

We now examine this in the high-frequency limit.  There we have
\begin{equation}\label{gameq}
\gamma _{aBC}=-\frac{i\omega}{2} \left( e^{-i\omega u}\phi 
  -e^{i\omega u}{\overline\psi}\right)m_a\omicron _B\omicron _C\, .
\end{equation}

\subsection{Response of the Fermi field}
  
Now let us consider the change $\delta\psi$ in the spinor field which accompanies the change in the metric.  For definiteness we take the spinor to satisfy the Dirac equation
\begin{equation}
  (-i\gamma ^a\nabla _a +m)\psi =0\, ,
\end{equation}
although we shall see soon that only the kinetic term will matter for us.
We first rewrite the equation in two-spinor form:
\begin{equation}
\left.\begin{matrix}
  -i\sqrt{2} \nabla ^{AA'}\psi _A +m\psi ^{A'}&=&0\\
  -i\sqrt{2} \nabla _{AA'}\psi ^{A'}+m\psi _A&=&0\end{matrix}
  \right\}\, . 
\end{equation}
Then the first-order perturbation is the system
\begin{equation}\label{psip}
\left.\begin{matrix}
-i\sqrt{2}\nabla ^{AA'}\delta\psi _A  +m\delta\psi ^{A'}&=&-i\sqrt{2}\gamma ^{AA'}{}_A{}^B\psi _B \\
-i\sqrt{2}\nabla _{AA'}\delta\psi ^{A'} +m\delta\psi _A&=&i\sqrt{2}\gamma _{AA'B'}{}^{A'}\psi ^{B'} \end{matrix}
 \right\}\, .
\end{equation}
Note that the right-hand sides \emph{vanish} to order $(-i\omega )$ (using eq. (\ref{gameq})).  In the high-fequency limit, we will have then
\begin{equation}
\delta\psi \sim O((-i\omega )^{-1})\, ,
\end{equation}
since in this limit the integration involved in solving eq. (\ref{psip}) dominates.

\subsection{Change in the stress--energy}

Now let us examine the change in the stress--energy.  The stress--energy for a Dirac field is
\begin{eqnarray}
T_{ab}&=&\frac{i}{2} {\tilde\psi}\gamma _b\nabla _a\psi
 -\frac{i}{2} \nabla _{(a}{\tilde\psi}\gamma_{b)}\psi\nonumber\\
 &=&\frac{i}{\sqrt{2}}\left\{
   {\overline\psi}_{B'}\nabla _a\psi _B +{\overline\psi}_B\nabla _a\psi _{B'}\right. \nonumber\\
   &&\left. -\nabla _a{\overline\psi}_{B'}\psi _B
  -\nabla _a{\overline\psi}_B\psi _{B'}\right\}\Bigr| _{{\rm sym}\, a\leftrightarrow b}\, ,
\end{eqnarray}
where in the last line the symmetrization is indicated at the very end, to avoid cluttering the equation.
There are two sorts of terms contributing to the change in this:  those which come from altering the derivative operators (which we write as $\delta _gT_{ab}$), and those which come from the perturbations of the Fermi field (written as $\delta _\psi T_{ab}$).  We have just seen that the Fermi field's change is of order $\delta\psi\sim O((i\omega )^{-1})$, so $\delta _\psi T_{ab}$, which involves first derivatives of $\delta\psi$, is of order $(i\omega )^0$, and will not contribute to $T^{(-1)}_{ab}$.  We therefore concentrate on $\delta _gT_{ab}$.

For the variation owing to the change in derivative operators, we have
\begin{widetext}
\begin{eqnarray}
\delta _g T_{ab} &=&\frac{i}{\sqrt{2}}\left\{
  -{\overline\psi}_{B'}\gamma _{aB}{}^C\psi _C
  -{\overline\psi}_B\gamma _{aB'}{}^{C'}\psi _{C'} 
 +\gamma _{aB'}{}^{C'}{\overline\psi}_{C'}\psi _B
  +\gamma _{aB}{}^C{\overline\psi}_C\psi _{B'}\right\}\Bigr| _{{\rm sym}\, a\leftrightarrow b}
  \nonumber\\
  &=&\frac{i}{\sqrt{2}}
  \left( \gamma_{aB}{}^C\epsilon _{B'}{}^{C'}
  -\gamma _{aB'}{}^{C'}\epsilon _B{}^C\right)  
  \left( -{\overline\psi}_{C'}\psi _C +{\overline\psi}_C\psi _{C'}\right)\Bigr| _{{\rm sym}\, a\leftrightarrow b}\nonumber\\
  &=&-\sqrt{2}\gamma^*_{ab}{}^c\left( {\overline\psi}_C\psi _{C'}-{\overline\psi}_{C'}\psi _C\right)\Bigr| _{{\rm sym}\, a\leftrightarrow b}\nonumber\\
  &=& \gamma ^*_{ab}{}^cj^5_c\Bigr| _{{\rm sym}\, a\leftrightarrow b}\, ,
\end{eqnarray}  
\end{widetext}
where we have put
\begin{equation}
\gamma _{aBB'CC'}=\gamma_{aBC}\epsilon _{B'C'}
  +\gamma _{aB'C'}\epsilon _{BC}\, ,
\end{equation}
and
\begin{equation}
  \gamma ^*_{abc}=\frac{1}{2}\epsilon _{bc}{}^{pq}\gamma _{apq}
\end{equation}
is its Hodge dual; in spinor form on the last indices
\begin{equation}
  \gamma ^*_{aBB'CC'}
  =-i\gamma_{aBC}\epsilon _{B'C'}
  +i\gamma _{aB'C'}\epsilon _{BC}\, .
\end{equation}

We recall that we will be interested in the component ${\overline m}^a{\overline m}^b\delta _gT_{ab}$.  Using eq. (\ref{gameq}), we have
\begin{equation}
{\overline m}^a{\overline m}^b\gamma ^*_{abc}
=\frac{\omega}{2}(e^{-i\omega u}\phi -e^{i\omega u}{\overline\psi})l_c\, ,
\end{equation}
and so
\begin{equation}
{\overline m}^a{\overline m}^b\delta _gT_{ab}
=\frac{\omega}{2}(e^{-i\omega u}\phi -e^{i\omega u}{\overline\psi})l^cj^5_c\, .
\end{equation}
Thus we have
\begin{eqnarray}
  {\overline m}^a{\overline m}^bT^{(-1)}_{ab}
    &=&+\frac{i}{2}l^aj^5_a\phi\\
m^am^bT^{(-1)}_{ab}&=&-\frac{i}{2}l^aj^5_a\psi\, .
\end{eqnarray}

\section{Nonperturbative approach}

There exist several approaches to the study of gravitational radiation.  I have emphasized so far a perturbative one, where the waves appear as small-amplitude high-frequency  oscillations on a background, because it
is common and also
most directly adapted to the cosmological questions at hand.  However, all such approaches suffer from the conceptual awkwardness that the background is
(despite its primacy in the analysis),
from a physical point of view, really a mathematical construct and not directly observable.  For this reason, I will here briefly indicate how the analysis here is related to the full, nonlinear theory.  

In the full theory, there is in general no unambiguous split of the metric into a background and waves.  One has rather a space--time, and we suppose that in the region of interest a standard null tetrad has been introduced, which, to good approximation, is adapted to the waves.  This means that we may at least approximately identify the wave-fronts as null hypersurfaces ruled by $l^a$, with $m^a$ a spacelike tangent; that is, we may say that waves are present if the 
curvature's components with respect to this tetrad have a suitable wavelike behavior.  (Most often, one considers an isolated system with Bondi--Sachs asymptotics and an associated tetrad; then one has a clean definition of the radiation in the limit of escape from the system.  However, as the point here is only to indicate the general way the chiral current enters into the full theory, there is no need to specialize to Bondi--Sachs space--times.)

In these circumstances, what happens is that each component of the stress--energy can be written as a sum of terms of two types:  those with derivative operators acting on components of the spinor $\psi$ with respect to a spin-frame associated with the tetrad, and those with the spinor components appearing only algebraically but multiplied by a connection coefficient.  The most important component for understanding the waves is
\begin{widetext}
\begin{eqnarray}\label{Tmmfull}
  T_{ab}m^am^b &=&i2^{-1/2} m^{BB'}\left\{
    {\overline\psi}_{B'}\eth \psi _B
    +{\overline\psi}_B m\eth \psi _{B'}
    -(\eth {\overline\psi}_{B'})\psi _B
    -(\eth {\overline\psi}_B) \psi _{B'}\right\}\\
      &=&i2^{-1/2}\left\{
    \iota ^{B'}{\overline\psi}_{B'}\eth\omicron ^B\psi _B
    +\omicron ^B{\overline\psi}_B\eth \iota ^{B'}\psi _{B'}
    -(\eth \iota ^{B'}{\overline\psi}_{B'})\omicron ^B\psi _B
    -(\eth\omicron ^B{\overline\psi}_B)\iota ^{B'}\psi _{B'}\right\}
    +i2^{-1} (\sigma n_a-{\overline\sigma}' l_a)j_5^a
 \, ,\nonumber
\end{eqnarray}    
\end{widetext}
where $\omicron ^A$, $\iota ^A$ form the spin-frame, the operator $\eth$ is a derivative in the $m^a$ direction, and $\sigma$, $\sigma '$ are the shears of the congruences formed by $l^a$ and $n^a$ \cite{PR1984}.  When waves associated with the $l^a$ congruence are present, we expect the last term, proportional to ${\overline\sigma}'l_aj_5^a$, to dominate. 

To study the effect systematically from this perspective, one would use the spin-coefficient formalism \cite{PR1984}.  I will here only note two of those equations which are relevant.  Taking the spin-frame to be propagated parallel along $l^a$, we have
\begin{equation}
\thorn\sigma ' = \rho\sigma ' +\rho ' \overline\sigma -4\pi Gm^am^bT_{ab}\, ,
\end{equation}
where $\thorn$ is a derivative in the $l^a$ direction and $\rho$, $\rho '$ are the convergences of $l^a$, $n^a$.  From this equation and eq. (\ref{Tmmfull}), we can see how $j_5^a$ contributes to the change of phase of $\sigma '$ along the geodesics ruled by $l^a$, and this may be a dominant effect in certain circumstances.
From
\begin{equation}
\thorn '\sigma ' -\eth'\kappa ' = (\rho '+{\overline\rho}')\sigma ' -{\overline\tau}\kappa ' +\Psi _4\, ,
\end{equation}
where $\thorn '$ is a derivative along $n^a$ (and $\eth '$ one along ${\overline m}^a$, and $\kappa '$ a measure of the acceleration of $n^a$, and $\tau$ of the rate of change of $l^a$ along $n^a$), we see
that the change of $\sigma '$ in the $n^a$ direction contributes to the curvature component $\Psi _4$, so it is a sort of first integral (with respect to a null coordinate $u$ with $l_a=\nabla u$) of the curvature's wave profile.

\bibliography{gyrotropypaperfinal.bbl}

\end{document}